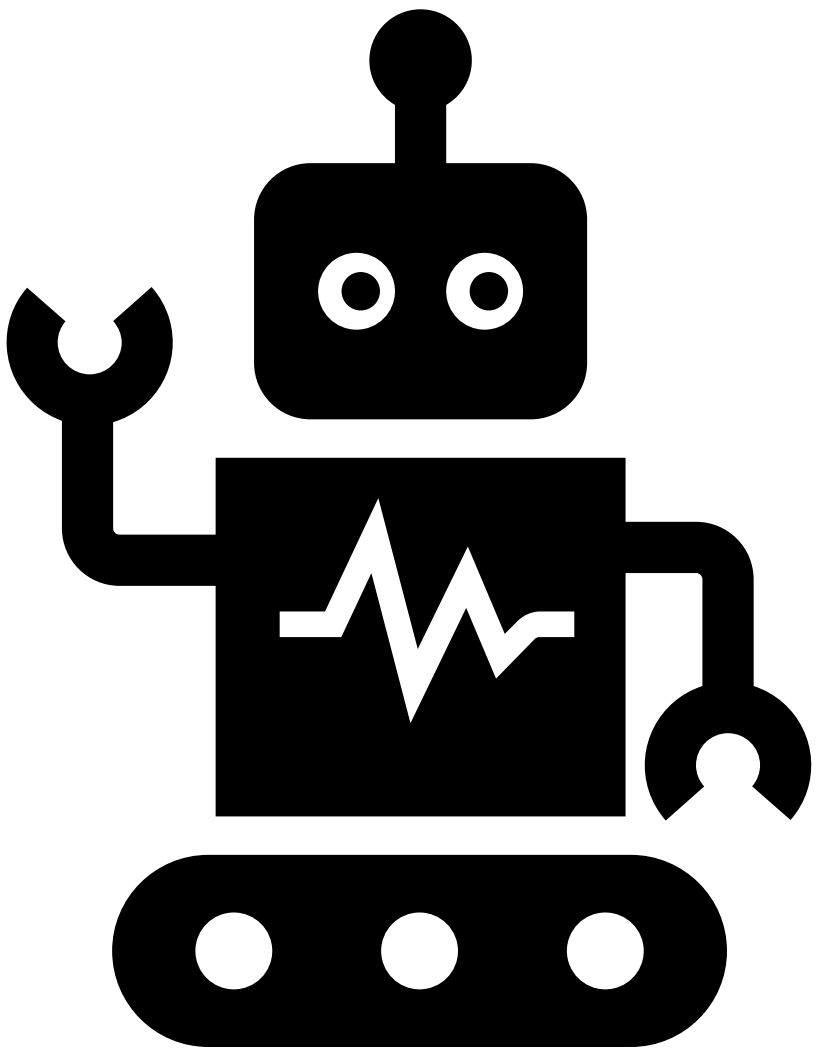

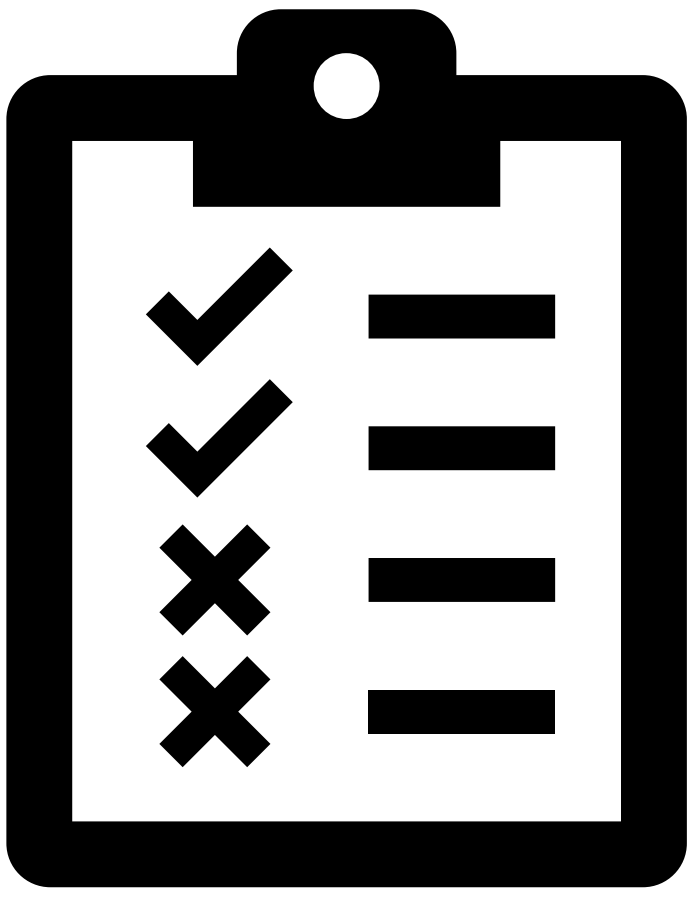

# Artificial intelligence technologies to support research assessment: A review

*Supplementary Report about the Responsible Use of Technology-Assisted Research Assessment*

**June 2022**

**Authors**

Kayvan Kousha and Mike Thelwall

Statistical Cybermetrics and Research Evaluation Group, University of Wolverhampton, UK.

# Table of Contents

# Executive Summary

This document provides literature review for ***The Responsible Use of Technology-Assisted Research Assessment*** project that was commissioned by the This study was funded by the four UK higher education funding bodies as part of the Future Research Assessment Programme (https://www.jisc.ac.uk/future-research-assessment-programme) to assess how technology, in the form of Artificial Intelligence (AI), can help research evaluation in the future, especially for the Research Excellence Framework (REF). Here, AI is essentially software that automates complex tasks.

The literature review identifies indicators that associate with higher impact or higher quality research from article text (e.g., titles, abstracts, lengths, cited references and readability) or metadata (e.g., the number of authors, international or domestic collaborations, journal impact factors and authors' h-index). This includes studies that used machine learning techniques to predict citation counts or quality scores for journal articles or conference papers. The literature review also includes evidence about the strength of association between bibliometric indicators and quality score rankings from previous UK Research Assessment Exercises (RAEs) and REFs in different subjects and years and similar evidence from other countries (e.g., Australia and Italy). In support of this, the document also surveys studies that used public datasets of citations, social media indictors or open review texts (e.g., Dimensions, OpenCitations, Altmetric.com and Publons) to help predict the scholarly impact of articles. The results of this part of the literature review were used to inform the experiments using machine learning to predict REF journal article quality scores, as reported in the AI experiments report for this project.

The literature review also covers technology to automate editorial processes, to provide quality control for papers and reviewers' suggestions, to match reviewers with articles, and to automatically categorise journal articles into fields. Bias and transparency in technology assisted assessment are also discussed.

**Recommendations:**

In addition to the analysis of inputs for the AI system to predict REF journal article scores, as discussed in the main report, the following recommendations are made based on the literature review.

1. Implement a system to recommend sub-panel members to review outputs. This would likely be based on the ORCIDs of sub-panel members matching their Scopus/Web of Science/Dimensions/etc. profiles, then using text mining to assess the similarity of their outputs with each sub-panel output to be assessed. The text mining might use article titles, abstracts, field classifications and references.
2. Build for the long-term implementation of quality control systems for academic articles by recommending that preprints of outputs for the next REF are saved in format suitable for text mining. Ideally, this would be in a markup format, such as XML, rather than PDF. This will also help longer-term AI systems for predicting REF journal article scores with article full text processing. At the end of the next REF, a future technology programme could then investigate the potential for full text mining for quality control purposes (e.g., checking statistics, plagiarism checks).
3. Build for the long-term exploitation of open peer review by, at the end of the next REF, calling for a review of current progress in the use of AI to exploit open peer review to assess article quality. Whilst open peer review should not be used as an input because it can be too easily exploited, investigations into its properties might shed useful light on aspects of quality identified by reviewers. Research into this is likely to occur over the next few years, and a review of it near the next REF might provide useful insights for both future AI and future human peer review guidelines for sub-panel members.

4. In the next REF, collate information on inter-reviewer agreement rates within sub-panels for outputs scored before cross-checking between reviewers. Use this to assess the human level agreement rates (for all output types) to use as a benchmark for score prediction AI systems.
5. In the tender for bibliometrics and AI for the next REF (if used), mention the importance of accurate classification for bibliometric indicators, including for the percentile system currently used.
6. Warn sub-panel members of the potential for small amounts of bias in the bibliometric data and AI (if used) and continue with the anti-bias warnings/training employed in REF2021.

# 1. Predicting journal article citation counts or quality from article text

Text mining approaches for predicting journal article citation counts or article quality have been applied to (a) the article title, abstract and keywords, (b) the article full text, or (c) the text of reviews of the article. Although not a heavily researched area, some studies have also investigated the readability of articles, often showing that more readable articles (or abstracts) tend to be more cited.

## 1.1. Article title and citation impact

Interesting or informative titles may attract the attention of other researchers, making the articles more likely to be read and then cited. Many studies have investigated the relationship between different characteristics of article titles (e.g., length, readability, and the presence of punctuation) and their subsequent citation counts in various subject areas.

### 1.1.1. Article title length and citation counts

Investigations of the relationship between citation counts and article title length, measured in words or characters, have generated mixed results for unknown reasons so there is not a simple and universal relationship. The following studies analysed relatively few articles from small sets of journals.

**Longer titles associate with more citations**: One simple study gathered the 25 most cited and 25 least cited articles from three medical journals (Lancet, BMJ and Journal of Clinical Pathology; total n=150) published in 2005, finding that number of title words positively correlates (rho=0397) with article citations (Jacques & Sebire, 2010). Similarly, a study of 9,031 articles published in 22 medical (e.g., Lancet and JAMA) and multidisciplinary (e.g., Science and Nature) journals in 2005 also found that articles with longer titles either as measured in characters (rho=0266) or in words (rho=o244) received more citations and this positive association was more common in high impact journals (7 of 8 journals) (Habibzadeh & Yadollahie, 2010).

**Insufficient evidence of an association between title length and citations**: In contrast to the above, a paper about 302 research articles published in the journal Addictive Behaviors in 2007 found no correlation between the number of title words and citation counts (Rostami, Mohammadpoorasl, & Hajizadeh, 2014). Similarly, no meaningful association was found between title length and citation counts for 1,825 articles published during 1990 to 2002 in five major marketing journals (Stremersch, Verniers & Verhoef, 2007). An investigation of 2,172 open access articles published in six PLoS (Public Library of Science) journals in 2007 also found no significant correlation between title length and citations (Jamali & Nikzad, 2011).

**Shorter titles associate with more citations**: In psychology, a study of 258 articles from 40 journals showed that articles with shorter titles received more citations (Subotic & Mukherjee, 2014). Similarly, an investigation of 423 research articles published in 2008 from BioMed Central (BMC) and Public Library of Science (PLoS) journals found a negative association (albeit very weak, r=-0.104) between title length (number of characters) and article citations (Paiva et al., 2012).

All the above studies investigated relatively few journals from a single year. The results - positive, negative or no association between title length and citation counts – do not reveal a simple pattern in terms of the disciplines covered (e.g., biomedical studies found both positive and negative associations) and hence it does not seem possible to generalise them into a general rule for the relationship between title length on article citations. Several large-scale studies have also investigated the relationship between article titles lengths and citation counts. These have different problems in that any differences found could be due to journal or topic style variations.

**Time differences in associations between title length and citations**: A study of 302,048 Web of Science economics articles from 1956–2012 found that correlations between title length and citation counts were negative between 1956 and 2000 but positive after 2000 (Guo et al. 2018). This result is partly corroborated by another large report on 140,000 highly cited Scopus papers from 2007-2013 (20,000 papers in each year). It found that highly cited articles with shorter titles tend to attract more citations (negative correlation with title length) (Letchford, Moat, & Preis, 2015). Supporting the correlation found in the second half of the Guo et al. investigation, a large study of 1.3 million articles published in 2012 found a weak positive correlation (rho=0.142) between the number of characters in title and citation counts (Haustein, Costas, & Larivière, 2015), although no other large-scale study seems to have reported changes over time in the relationship between title length and citations. Thus, it seems that there may be a general trend for articles with longer titles to be slightly more cited overall now, reversing an earlier trend. This interdisciplinary finding could be a second order effect of disciplinary differences in title lengths and citation rates, however.

**Disciplinary differences in associations between title length and citations**: A study of articles published in journals with the highest impact factors during 1996-2005 found that articles with longer titles received fewer citations in both Sociology (n=2,016; r=-0.046) and Applied Physics (n=23,676; r=-0.089), but the reverse in General Medicine (n= 6,957; r=-0.166) (Van Wesel, Wyatt, & ten Haaf, 2014). For Web of science articles in Biology and Biochemistry (n=16,058) and Social Sciences (n=15,932) from 2000-2009, title length associates with more citations in both subject areas according to both regression models and Spearman correlations (rho=0.021 and 0.014 respectively) but no association was found in Chemistry (n=16,378) (Didegah & Thelwall, 2013b).

**Journal impact differences in associations between title length and citations**: One of the largest studies investigating the association between article length and citations used 4.3 million papers in articles published 1995-2004 in 1500 large journals, finding that for highly cited journals, shorter titles tend to be more cited, whereas for the remaining journals, longer titles tend to be more cited (Sienkiewicz & Altmann, 2016). The former result may be due to strict title length restrictions in the most highly cited journals within the highly cited set.

Despite all the above studies, it is still not fully clear if (a) shorter titles are more common in high impact journals than in other journals or (b) if longer (or shorter) title lengths have a citation advantage in some fields because they increase readership and therefore are more likely to be subsequently cited. The contradictory findings might be related to other factors related to the studied data sets such as a focus on highly cited articles or journals (Subotic & Mukherjee, 2014; Sienkiewicz & Altmann, 2016), different publication years examined because article title lengths have increased over time (Lewison & Hartley, 2005; Guo et al. 2015; Gnewuch & Wohlrabe, 2017), differing document types (e.g., see Soler, 2007) or other factors, such as changing editorial policies regarding title lengths.

### 1.1.2. Non-alphanumeric title characters and citation impact

There is evidence that non-alphanumeric characters in articles can have an impact on citation counts, presumably because they reflect successful article styles, such as asking a question.

**Colons and hyphens in titles associate with more citations**: A colon or a hyphen in an article title may help to make it more readable or may help to express more complex information. These may associate with more citations. For instance, an early study of 150 articles from three medical journals (Lancet, BMJ and Journal of Clinical Pathology; total n=150) showed that colons was significantly more frequent in highly cited articles (Jacques & Sebire, 2010).

**Colons and hyphens in titles associate with less citations**: Two small studies (n=423 and n=2,172) found that biomedical article titles with colons or hyphens in their titles received fewer citations than titles that did not have these characters (Paiva et al., 2012; Jamali & Nikzad, 2011, respectively). The difference from the above set is presumably the wider collection of articles analysed.

**Question marks in titles associate with more citations**: A paper about 312,879 Web of Science articles in economics published between 1980 and 2015 in 430 journals found non-alphanumeric characters in article titles associated with increased their citation impact, with question marks having the strongest apparent effect: 1.64 extra citations (Gnewuch & Wohlrabe, 2017). Similarly, a recent large-scale study of about 2 million Web of Science journal articles and conference proceedings papers (1945 to 2014) in Computer Science also found that articles with titles ending with a question mark (n=5,682) received 16% more citations than articles not asking questions (n=957,837) (Fiala, Král, & Dostal, 2021).

**Non-alphanumeric characters in titles associate with more citations**: A large-scale study of 5% of all Web of Science articles from 1999-2008 (n=642,807) found that 68% had at least one out of 29 non-alphanumeric characters in their titles, with hyphens, colons, and commas being the most common. In general, articles with non-alphanumeric characters in their titles had higher field-normalised citation impact than titles with only alphanumeric characters. However, there were disciplinary differences, and this association was positive in Clinical Medicine, negative in Biological Sciences, and no significant association was found in Agriculture and Food Science (Buter and van Raan 2011).

### 1.1.3. Other characteristics in titles and citation counts

Several other studies in different fields have suggested that using humour (Sagi & Yechiam 2008), unusual words that are rarely used in other titles in the same fields (Thelwall, 2017c), stylistic cues such as metaphor or alliteration (Keating et al., 2022) and quoted or direct speech (Pearson, 2021) in research article titles can have a significant negative association with citation counts.

## 1.2. Article length and citation counts

A hypothesis tested by several studies is that longer papers tend to attract more citations. The assumption is that longer articles provide more information or analysis about the investigated topic and therefore contain more citeable content. Although most findings have been positive, some prestigious journals require short articles and so the relationship is not universal. All studies reviewed here measured article length using the total number of pages. Nevertheless, page counts may depend on page layouts and printing formats (e.g., columns and font size), which vary between journals and over time. Although word counts could be a better indicator to measure article length, the article full text is needed, and this could be more difficult for a large-scale analysis and when using PDF version of articles.

**Longer articles are more cited**: An investigation of 13,125 articles in immunology and 17,083 articles in Surgery found a weak but significant correlation (rho= 0.286 and 0.335 respectively) between page counts and citations per article (Weale, Bailey, & Lear, 2004). A paper about 26,088 articles published in 32 ecology Web of Science indexed journals (2009-2012) found that both article page and reference counts positively correlated with citation counts, suggesting that longer articles attract more citations (Fox, Paine, & Sauterey, 2016). Similarly, a large-scale study of research articles published in The New England Journal of Medicine (n=27,305), The Journal of the American Medical Association (n=42,733) and The Lancet (n=65,525) from their formation to 2016 found that the length of articles (as measured by the number of pages) published by these journals had increased over time and longer articles tended to receive more citations (Lyu & Wolfram, 2018). A multidisciplinary study of articles from Sociology (n=2,016), Applied Physics (n=23,676) and General Medicine (n=6,957) found significant positive correlations between page counts and citation counts (Van Wesel, Wyatt, & ten Haaf, 2014). Similarly, another multidisciplinary study of Web of Science articles found that the size of the paper (number of pages) positively associated with citation counts for articles in Biology & Biochemistry (n=44,248; $R^2$=0.666), Chemistry (n=97,177; $R^2$=0.355), Mathematics (n=20,127; $R^2$=0.0864) and Physics (n=64,614; $R^2$=0.615) (Vieira & Gomes, 2010). A recent report about 1,561 articles published in five economics journals during 2010-2014 found a 1% increase in article page count associated with 0.56% more Google Scholar citations (Hasan & Breunig, 2021). A positive association, albeit weak (rho=0194), between article page counts and citation counts has been also confirmed in a large study of 1.3 million articles published in 2012 (Haustein, Costas, & Larivière, 2015). Other relatively small-scale studies (see also the summary section) have also reported significant low to medium positive correlations between article lengths (all measured by the number of pages) and citation counts in psychology (Haslam et al., 2008, n=308), psychiatry (Hafeez, Jalal, & Khosa, 2019, n=545), medicine (Falagas et al. 2013, n=196), management (Mingers & Xu 2010, n= 696) and social sciences (Hodge et al., 2017, n= 3,066). Finally, a random effects meta-analysis study found a moderate positive overall correlation (r=0.310) between article length and citations (Xie et al, 2019).

**Longer articles are not more cited**: One investigation of Biology and Biochemistry (n=16,058) and Social Sciences (n=15,932) articles from 2000-2009 found no significant associations between article page and citation counts (Didegah & Thelwall, 2013b).

## 1.3. Abstract length and citation counts

Abstracts have become near-universal for journal articles over the past half century (Thelwall & Sud, 2022). They help potential readers to understand the topic and results of an article efficiently before they read the full article. Presumably, informative and reliable abstracts can help relevant research to be quickly identified. Many relevant studies have investigated abstract length, structure or readability. This section focuses on abstract length.

**Longer abstracts tend to be more cited**: One of the early studies of The Lancet articles during 1997–1999 found that the median number of words in abstracts was 2.5-7 higher in articles with the most citations compared with articles with the least citations (Kostoff, 2007). Such large differences in abstract lengths seem unlikely in the current era of greater journal format standardisation, however. A large multidisciplinary study of one million abstracts from eight subject areas showed that longer abstracts and more sentences in abstracts associated with more citations in all fields, although in Mathematics and Physics abstracts with shorter sentences attracted more citations (Weinberger, Evans, & Allesina, 2015). Similarly, longer abstracts significantly correlate with more citations in Biology and Biochemistry (n=16,058) and Social Sciences (n=15,932) and Chemistry (n=16,378), although this association is very weak, rho=0.153, 0.122 and 0.148 respectively (Didegah & Thelwall, 2013b). A very large study of 4.3 million papers from over 1500 journals also showed that the length of abstracts positively correlated with citation counts in nearly all of the journals (Sienkiewicz & Altmann, 2016). Positive associations between citation counts and abstract length might be a statistical side-effect of a minority of minor articles having very short abstracts, however. These could be errors (e.g., corrections published as articles), or short articles or comments with a few summary sentences instead of a detailed abstract.

**Shorter abstracts tend to be more cited**: In contrast to the above studies, another large-scale investigation of 300,000 highly cited articles between 1999 and 2008 (30,000 papers per year) found that articles with shorter abstracts received more citations, with this pattern being found at the journal level. The authors argued that some high impact journal might limit abstract size (e.g., 125 words for the journal Science) and hence the selection of highly cited papers could influence the result (Letchford, Preis, & Moat, 2016).

**Keyword repetition associates with more citations**: One study introduced an abstract ratio indicator (the sum of repetition of keywords in abstract divided by abstract length), finding that it statistically correlates with citation counts for 5,875 articles in Education (Sohrabi & Iraj, 2017, p. 250). Keyword repetition may suggest a narrower focus for a study or an emphasis on a key message. Another study tested five keyword popularity features, finding that keyword popularities can more effectively predict highly cited papers (n=746 articles from 46 journals in marketing and MIS) than can author (author's h-index, publications or citations) and journal (e.g., journal impact factor and SCImago) features (Hu et al., 2020).

## 1.4. Article readability (abstract or full text) and citation counts

Several studies have investigated whether more readable abstracts associate with higher citation rates, mostly finding the opposite. There are many ways of measuring readability, such as the relative frequency of rare or long words, with no measure being recognised as the best.

**Articles with more readable abstracts are less cited**: Using Flesch Readability scores to assess the readability of abstracts, a study of 264,156 articles from five American universities during 2000–2009 found that abstracts with more difficult language (e.g., with scientific or technical terms) were cited more than articles with easier abstracts to read, ranging from r=–0.469 to –0.821 (Gazni, 2011). Abstract readability, as measured by the Flesch reading ease score, was found to significantly correlate, albeit weak (rho=0.073), with fewer citations in Biology & Biochemistry and no significant relationship was found between the readability of abstracts and citation counts in Social Sciences and Chemistry (Didegah & Thelwall, 2013b). Another study using the full text of 1,825 articles from five major marketing journals during 1990 to 2002 also found that Flesch reading ease scores negatively correlated (r=0.02) with citations (Stremersch, Verniers & Verhoef, 2007). Using the Coh-Metrix analysis tool to assess the difficulty of written text based on computing computational cohesion and coherence metrics, a more recent study of 10,000 highly cited and 10,000 uncited English language research articles published during 2008-2017 across 22 subject areas found that abstracts of highly cited articles contained more complex, difficult and professional terms. The highly cited articles also had more adjectives, adverbs, conjunctions and personal pronouns and longer sentences, making them less readable compared with abstracts of uncited articles (Hu, Wang, & Deng, 2021). Using the Python package Readability 0.3.1, a recently published study of 135,502 abstracts from research articles on 12 emerging technologies (e.g., artificial intelligence, big data, and virtual reality) by the end of 2020 found that abstracts had become more difficult to understand over time across all emerging technology topics and articles with more complex abstracts received fewer citations or remained uncited (Ante, 2022). A large-scale paper about over 4.3 million papers from over 1500 journals also found that the text complexity of abstracts positively correlated with citation counts (Sienkiewicz & Altmann, 2016).

**Articles with more readable abstracts are more cited**: In contrast to the above investigations, a study of 3,229 articles published in the journal Economics Letters during 2003–2012 found a positive correlation between Flesch Reading Ease Score of abstracts and citation counts, suggesting that clearer writing can potentially enhance the impact of economic research (Dowling, Hammami, & Zreik, 2018). Economics might be an exception to the general rule in science in this regard.

**Articles with more readable abstracts are similarly cited**: Analysing the full text of articles in Biology (n=36,400) and Psychology (n=1,797), a study found no practical significant correlation between the linguistic complexity of scientific articles and their citation impact (Lu et al., 2019).

**Articles with structured abstracts are more cited**: Structured abstracts (usually with sub-headings such as background, method, aim or results) could be easier to read than traditional abstracts (Hartley & Sydes, 1997). In support of this, a small study found weak but positive correlation (rho=0.03) between structured abstracts and citation counts (n=545) from six high impact psychiatry journals (Hafeez, Jalal, & Khosa, 2019).

**Articles with structured abstracts are less cited**: Structured abstracts negatively correlated with citation counts in another study of 757 clinical articles from medical journals (Lokker et al., 2008).

## 1.5. Cited references and citation impact

Citing more references in articles may be one of the factors influencing the scholarly impact of articles and make them more visible to researchers using citation tracing in citation databases such as the Web of Science, Scopus and Google Scholar. Longer articles with more content may also tend to have more references and be cited more. To investigate this, many studies have assessed features of cited references and their association with citation metrics.

**Articles with more references are more cited**: A large study of 226,166 Web of Science articles from 2004 in Biology & Biochemistry (n=44,248; $R^2$=0.917), Chemistry (n=97,177; $R^2$=0.858), Mathematics (n=20,127; $R^2$=0.799) and Physics (n=64,614; $R^2$=0.846) found that reference counts correlate positively with citation counts (Vieira & Gomes, 2010). Another more recent study of articles published in seventeen ecological journals between 1997 and 2017 (n=50,878) confirmed that articles with more references are more cited (Mammola et al., 2021). An analysis of 757 clinical articles from 105 medical journals published from January to June 2005, finding that reference counts and other factors (e.g., number of authors; indexing in numerous databases) associate with increased citations (Lokker et al., 2008). Other relatively small-scale studies in Artificial Intelligence (Xiao & Jiang, 2020, n=9,458), psychology (Haslam et al., 2008, n=308), psychiatry (Hafeez, Jalal, & Khosa, 2019, n=545), library and information science (Yu et al., 2014, n=1,025) and management (Antonakis et al., 2014, n=776) also reported low to medium correlations between reference counts and citation counts for articles (see also Table 1 ).

**Articles with more recent references are more cited**: Using quantile regression and normalisation of all variables, a very large study of 955,663 Web of Science research articles published in 2009 found that more references and more recent references in academic articles positively correlate with their citation impact (Ahlgren, Colliander & Sjögårde, 2018). A report on 1,395 articles published in 2000 across five science and one engineering subjects (n=230–240 in each field) found that the recency of cited references, as measured by the Price index (percentage of the references within five years before the publication year of the article), had the strongest association with the citation counts, followed by the number of references (Onodera & Yoshikane, 2015).

**Articles with more high impact references are more cited**: There is substantial evidence that articles with high impact cited references tend to be more cited. For instance, a very large-scale study of 780,049 articles from The Science Citation Index Expanded and The Social Sciences Citation Index 2002-2003 found a significant correlation between the citation counts of articles and the citation impact of their cited references (the number of times the article's references were cited), suggesting that an academic article with highly cited references is more likely to become more cited, although there were disciplinary differences (Boyack & Klavans, 2005). Later studies confirmed that reference impact could be an indicator of future citations. A study of over 1.6 million articles from Scopus in 2007 found that both the number and citation impact of the cited references of articles correlated with their citation counts (Lancho-Barrantes, Guerrero-Bote, & Moya-Anegon, 2010) and another paper about 1,765 chemical papers calculated the h-index for the cited references of each article, finding a statistically significant correlation with citation counts (Bornmann, Schier, Marx, & Daniel, 2012). From a related perspective, an investigation of 7,749 articles published in 105 journals related to Internet studies found that the authoritativeness of the cited references (the proportion of highly cited references among total cited references in the topic) had a significant positive correlation with citation counts ($\gamma = 0.988$, $p < 0.001$) (Peng & Zhu, 2012). For 50,162 articles in nanoscience and nanotechnology journals published 2007-2009, the number of cited references and their average citation impact significantly correlated with the articles' citation counts (Didegah & Thelwall, 2013a). Similar associations were found for articles in Biology & Biochemistry (n=16,058), Social Sciences (n=15,932) and Chemistry (n=16,378) published during 2000-2009 (Didegah & Thelwall, 2013b).

**Articles with more international references are more cited**: There is evidence that the internationality of cited references significantly correlates with citation counts from a study of 50,162 articles in nanoscience and nanotechnology from 2007-2009 (Didegah & Thelwall, 2013a).

## 1.6. Other article features and citation counts

This section reviews article features that have not been extensively investigated for associations with citation counts.

**Images in articles**: Analysing over 4.8 million figures from 650,000 PubMed articles, higher-impact articles, as measured by article-level Eigenfactor, had more diagrams per page and a higher proportion of diagrams but a lower proportion of photos (Lee, West, & Howe, 2017).

**Review articles are more cited**: Review articles tend to be more cited than other research articles, although there are some disciplinary differences (e.g., Aksnes, 2006; Colebunders, Kenyon, & Rousseau, 2014; for a review see Blümel & Schniedermann, 2020). For instance, a very large-scale study of 14.2 million records from Science Citation Index Expanded database during 2000–2015 across 35 science subject areas found that reviews received 1.3 to 6.7 times more citations than standard research articles, depending on the subject area (e.g., much higher in Engineering Electrical Electronic, Chemistry Multidisciplinary, Physics Applied and Materials Science than Oncology, Radiology Nuclear Medicine, Surgery, and Mathematics (Miranda & Garcia-Carpintero, 2018).

## Table 1. Studies associating journal article citation counts with information extracted from article text.

| Factors | Study | No. of articles | Dataset | Subject | Association with citation | Main result |
|---|---|---|---|---|---|---|
| The length of article title and citation counts | Jacques & Sebire, 2010 | 150 | 25 most cited and 25 least cited articles form the Lancet, BMJ J Clin. Pathology in 2005. | Medical Sci. | Positive | Words in titles significantly correlated with citation counts (rho=0.397, p=0.004). |
| | Paiva et al., 2012 | 423 | Articles from the BioMed Central and PLoS journals (n=19) published in 2008. | Medical Sci. Biomedical | Negative | Articles with shorter titles had more citations than those with longer titles (r=-0.104, p=0.032). The number of title characters was a statistically significant predictor of citation counts (F=7.581, p=0.001). |
| | Habibzadeh & Yadollahie, 2010 | 9,031 | Articles from 22 medical and multidisciplinary journals in 2005 with high, medium, and low journal impact factors. | Medical Sci. Biomedical Multidiscip. | Positive | Articles with longer titles received more citations either measured in character (rho=0.266) or in words (rho=0.244) both at p<0.001. However, this association was more common in high impact journals (7 of 8) than other journals (2 of 14). |
| | Rostami et al., 2014 | 302 | Articles published in the journal Addictive Behaviors in 2007. | Psychology | No association | The number of words in the title was not correlated with citation counts. |
| | Stremersch et al., 2007 | 1,825 | Articles from five major marketing journals during 1990 to 2002 | Marketing | No association | The number of words in the title was not correlated with citation counts. |
| | Jamali & Nikzad, 2011 | 2,172 | Articles from six PLoS (Public Library of Science) journals in 2007. | Biomedical (excl. PLoS One,) | No association | No significant correlation found between title lengths and citation counts. |
| | Guo et al. 2018 | 302,048 | Web of Science articles in Economics during 1956–2012. | Economics | Negative association between 1956-2000 but became positive between 2001–2012 | Correlation between title length and the citation counts was significantly negative between 1970s to 1990s (Coefficients ranging from -0.00417 to -0.00818, respectively), but becomes positive after 2000s (0.00596). |
| | Letchford et. al., 2015 | 140,000 | Highly cited Scopus articles during 2007-2013. | General | Negative | Highly cited articles with shorter titles receive more citations (Kendall's τ coefficient=−0.042, p <0.001) |
| | Van Wesel et al., 2014 | 2,016 (Sociology) 6,957 (Gen. Med) 23,676 (Physics) | Articles published in journals with the highest impact factor during 1996-2005. | Sociology Gen. Med. Physics | Negative association in Sociology and Physics, but positive in Gen. Med. | Articles with shorter titles received more citations in both Sociology (r=-0.046) and Applied Physics (r=-0.089), but in Medicine longer titles had more citations (r=0.166) all at p= 0.01. |
| | Didegah & Thelwall, 2013b | 16,058 (Bio. Sci.) 15,932 (Social Sci.) 16,378 (Chem.) | Web of Science indexed articles during 2000-2009. | Biology & Biochem. Social Sci. Chemistry | Negative association in Biology & Biochem. and | Articles with shorter titles had more citation in Biology & Biochemistry (rho=0.021) and Social Sciences (rho= 0.014) and there was no association in Chemistry. |
| | | | | | Social Sci., but no association in Chemistry | |
| | Gnewuch & Wohlrabe, 2017 | 312,879 | Articles in economics published between 1980 and 2015 in 430 journals. | Economics | Negative | Shorter article titles associate with increased citation counts (coefficient= -0.15). |
| | Haustein et al., 2015 | 1.3 million | Articles published in 2012 across different fields. | - | Positive | Articles with more characters in title received more citations (rho=0.142). |
| | Sienkiewicz & Altmann, 2016 | 4.3 million | Articles from over 1,500 journals during 1995–2004. | General | Negative or positive depending on articles (highly cited and normal) | Shorter titles of highly cited articles positively correlated with citation counts, whereas for normal papers, longer article titles received more citations. |
| The non-alphanumeric character in title and citation impact | Jacques & Sebire, 2010 | 150 | 25 most cited and 25 least cited articles form the Lancet, BMJ J Clin. Pathology in 2005. | Medical Sci. | Positive | Colons were significantly more common in highly cited articles compared with least cited articles (Z=2.3, P=0.02). |
| | Paiva et al., 2012 | 423 | Articles from the BioMed Central and PLoS journals (n=19) published in 2008. | Biomedical | Positive | Article titles with two components separated by a colon or a hyphen had fewer citations compared with titles that did not have these characters (p =0.004). |
| | Jamali & Nikzad, 2011 | 2,172 | Articles from six PLoS (Public Library of Science) journals in 2007. | Biomedical (excl. PLoS One,) | Positive | Articles with colon in their titles received fewer citations (median = 9) compared to titles without colons (median= 12; p=0.012). |
| | Buter and van Raan 2011 | 642,807 | About 5% of Web of Science articles during 1999-2008 across different subjects. | Multidisciplinary | Positive | In general, articles with non-alphanumeric characters in their titles had higher citation impact than titles with only alphanumeric characters, although there were disciplinary differences. |
| | Gnewuch & Wohlrabe, 2017 | 312,879 | Articles in economics published between 1980 and 2015 in 430 journals. | Economics | Positive | A non-alphanumeric character in article title associates with higher citation impact (coefficient=0.47). Question marks had the greatest association, increasing the citation count by 1.64; 0.90 for colons. |
| | Fiala et al., 2021 | 1,922,652 | Articles and conference proceedings papers published during 1945 to 2014. | Computer Science | Positive | Citation counts per article asking questions (n=5682) was 16% higher than for other papers (n=957,837) and this difference was statistically significant at the 0.001 level. |
| The Impact of article length on citations | Weale et al., 2004 | 13,125 (Immun.) 17,083 (Surgery) | Articles in immunology and Surgery. | Immunology and Surgery | Positive | A weak but significant correlation (rho= 0.286 and 0.335 respectively) between the number of pages and citations per article in immunology and Surgery. |
| | Fox et al., 2016 | 26,088 | Articles published in 32 Ecology Web of Science indexed journals during 2009-2012. | Ecology | Positive | Longer articles in Ecology tended to be cited more (r= 0.147), although this association varied among journals based on manuscript length in author guidelines. |
| | Lyu & Wolfram, 2018 | 27,305 (NEJM) 42,733 (JAMA) 65,525 (Lancet) | Articles from three medical journals (New England Journal of Medicine, JAMA and Lancet) since their creation up to 2016. | Medical Sci. | Positive | The length of medical articles in the studied journals had increased over time and, on average, longer articles received more citations than shorter articles in all three journals. |
| | Van Wesel et al., 2014 | 2,016 (Sociology) 6,957 (Gen. Med) 23,676 (Physics) | Articles published in journals with the highest impact factor during 1996-2005. | Sociology Gen. Med. Physics | Positive | Number of article pages associates with citation counts in Sociology (r=0.122) and Physics (r=0.033), and strongly in General Medicine (r=0. 435), all significant with p<0.01. |
| | Vieira & Gomes, 2010 | 44,248 (Bio.) 97,177 (Chem.) 20,127 (Maths) 64,614 (Physics) | Web of Science articles in four science fields in 2004. | Bio. Biochem. Chemistry Maths Physics | Positive | Article length (page count) associates with higher citation counts for articles in Bio. & Biochem. ($R^2$= 0.666), Chemistry ($R^2$=0.355), Mathematics ($R^2$=0.864) and Physics ($R^2$=0.615) with citation enhancements of 50%, 30%, 62% and 37% respectively. |
| | Hasan & Breunig, 2021 | 1,561 | Articles published in five Economics journals during 2010-2014. | Economics | Positive | A 1% increase in article size predicts an increase in Google Scholar citations by 0.56%. |
| | Haslam et al., 2008 | 308 | Articles published in top five economics journals between 2010 and 2014. | Psychology | Positive | Longer articles tended to receive more citations (overall multiple regression analysis = 0.21 at p < 0.001). |
| | Hafeez et al., 2019 | 545 | Articles in 2007 from six major psychiatry journals in 2007. | Psychiatry | Positive | Number of pages of articles significantly correlated with citation counts (rho=0.15). |
| | Falagas et al. 2013 | 196 | Articles from five journals with highest impact factors in General Medicine in 2016. | Medical Sci. | Positive | Article length (number of pages) independently predicted the number of future citations (r=0.700). |
| | Hodge et al., 2017 | 3,066 | Articles from 18 Social Work journals published during 2005 to 2009. | Social Sci. | Positive | Longer articles tend to receive more citations and every additional page was associated with almost 4% more citations (rho=0.09). |
| | Didegah & Thelwall, 2013b | 16,058 (Bio. Sci.) 15,932 (Social Sci.) 16,378 (Chem.) | Web of Science indexed articles during 2000-2009. | Biology & Biochem. Social Sci. Chemistry | No association | No significant associations found between article length and citations. |
| | Xie et al., 2019 | 1,548,088 (meta-analysis) | A meta-analysis of 18 relevant studies. | Different subjects | Positive | Meta-regression analysis of relevant studies found a moderate, positive correlation between article length and citations (r=0.310). |
| | Mingers & Xu 2010 | 696 | Articles published in six Management journals in 1990. | Management | Positive | Longer article attracted more citations (0.277). |
| | Haustein et al., 2015 | 1.3 million | Articles published in 2012 across different fields. | Multidisciplinary | Positive | Article length significantly correlated with citation counts (rho=0.194). |
| The length of abstract and citation counts | Weinberger et al., 2015 | one million | Abstracts from articles with abstracts from eight science subjects within 17 years. | Science | Negative | Shorter abstracts associated with decreased citation impact for articles in all fields. |
| | Didegah & Thelwall, 2013b | 16,058 (Bio. Sci.) 15,932 (Social Sci.) 16,378 (Chem.) | Abstracts from Web of Science indexed articles during 2000-2009. | Biology & Biochem. Social Sci. Chemistry | Positive | Abstract length significantly correlated with increased citations in all fields, but this association found to be weak in Social Science (rho =0.122) in Chemistry (rs=0.148) and in Biology & Biochemistry (rho=0.153). |
| | Letchford et al., 2016 | 300,000 | Abstracts from most highly cited articles between 1999 and 2008. | General | Negative | Shorter abstracts receive slightly more citations at the journal level. Adding a 5-letter word decreases the predicted number of citations by 0.02%. |
| | Van Wesel et al., 2014 | 2,016 (Sociology)<br>6,957 (Gen. Med)<br>23,676 (Physics) | Abstracts from articles published in journals with the highest impact factor during 1996-2005. | Sociology<br>Gen. Med.<br>Physics | Positive | The length of the abstract (as measured by numbers of sentences), correlated positively with citations in both General & Internal Medicine (r=0.314) and Applied Physics (0.049), but not in Sociology. |
| | Sohrabi & Iraj, 2017 | 5,875 | Abstracts from articles in Education subject areas. | Education | Positive | Both abstract ratios (logistic regression= 5.216) and weight (logistic regression= 3.58) were significant variables in predicting future citations. |
| | Hafeez et al., 2019 | 545 | Abstracts from articles form six high impact psychiatry journals in 2007. | Psychiatry | Positive | Structured and longer abstracts tended to receive more citations. Both abstract character count (rho=0.22) and word count (rho=0.17) correlated positively with citation counts. |
| | Sienkiewicz & Altmann, 2016 | 4.3 million | Articles from over 1500 journals during 1995–2004. | Multidisciplinary | Positive | The number of words in the abstract positively correlated with citation counts in almost all studied journals. |
| Article readability and citation impact | Gazni, 2011 | 264,156 | Abstracts from articles from five American universities during 2000–2009. | General | Negative | Abstracts with more difficult language (as Flesch reading ease score) tended to attract more citations than abstracts with easier language. The Pearson correlations between citation per paper and Flesch score ranged from –0.469 to –0.821 all significant with p<0.01. |
| | Didegah & Thelwall, 2013b | 16,058 (Bio. Sci.)<br>15,932 (Social Sci.)<br>16,378 (Chem.) | Abstracts of Web of Science indexed articles during 2000-2009. | Biology & Biochem.<br>Social Sci.<br>Chemistry | Negative association in Bio. & Biochem.<br>No association in Social Sciences and Chemistry | Abstract readability (Flesch reading ease score) correlated significantly with decreased citations in Biology & Biochemistry (rho=−0.073), but no significant relationship was found in Social Sciences and Chemistry. |
| | Stremersch et al., 2007 | 1,825 | Full texts of articles from five major marketing journals during 1990 to 2002. | Marketing | Negative | Readability of articles (Flesch reading ease score) negatively associates with citations and more difficult texts tended to attract more citations (r=-0.02, p < 0.01). |
| | Hu et al., 2021 | 20,000 | Abstracts of 10,000 highly cited and 10,000 uncited articles published during 2008-2017. | General | Negative | Abstracts of highly cited articles tended to have more complex, difficult and professional terms than abstracts of uncited articles and this difference was significant with p < 0.01. |
| | Lu et al., 2019 | 36,400<br>1,797 | Full text of articles. | Biology<br>Psychology | No association | No practical significant relationship between linguistic complexity and citation impact. |
| | Ante, 2022 | 135,502 | Abstracts from articles on 12 emerging technologies subjects by the end of 2020. | Emerging technologies | Negative | The abstracts of the top 10% and 1% of the most frequently cited articles were significantly less readable and zero-cited articles on average were almost always easier to read. |
| | Hafeez et al., 2019 | 545 | Articles form six high impact psychiatry journals in 2007. | Psychiatry | Positive | Articles with structured abstracts associate with higher citation counts (rho= 0.03). |
| | Dowling, 2018 | 3,229 | Articles published in the journal Economics Letters during 2003–2012. | Economics | Positive | A positive correlation between Flesch Reading Ease Score of abstracts and citation counts. |
| | Sienkiewicz & Altmann, 2016 | 4.3 million | Articles from over 1500 journals during 1995–2004. | Multidisc. | Negative | The text complexity of abstracts positively correlated with citation counts. |
| The relationship between features of cited references and citation impact | Hafeez et al., 2019 | 545 | Articles form six high impact psychiatry journals in 2007. | Psychiatry | Positive | The number of references significantly corelated with citations (rho=0.2), |
| | Vieira & Gomes, 2010 | 44,248 (Bio & Biochem.)<br>97,177 (Chemistry)<br>20,127 (Maths)<br>64,614 (Physics) | Articles in four science fields in 2004. | Bio. & Biochem.<br>Chemistry<br>Maths<br>Physics | Positive | The number of references significantly predicted the citation counts of articles in Bio. & Biochem. ($R^2$= 0.917), Chemistry ($R^2$=0.858), Mathematics ($R^2$=0.799) and Physics ($R^2$=0.846) with citation enhancement 69%, 60%, 72% and 58% respectively. |
| | Ahlgren et al., 2018 | 955,663 | Articles published in 2009. | General | Positive | Using quantile regression, reference counts and more recent references in academic articles positively correlate with their citation impact. |
| | Didegah & Thelwall, 2013a | 50,162 | Articles in nanoscience and nanotechnology journals published between 2007-2009. | nanoscience and nanotechnology | Positive | Number, internationality, and impact of cited references predict increased citations. A one standard deviation increase in these three factors predicted a 19.2%, 17.3% and 35% increase in the citation counts, respectively. |
| | Didegah & Thelwall, 2013b | 16,058 (Bio. Sci.)<br>15,932 (Social Sci.)<br>16,378 (Chem.) | Articles during 2000-2009. | Biology & Biochem.<br>Social Sci.<br>Chemistry | Positive | The number and impact of cited references significantly correlated with article citations in Biology & and Biochemistry (rho= 0.265 and 0.416 respectively), Social Science (rho= 0.104 and 0.302) and Chemistry (rho= 0.304, 0.359). |
| | Mammola et al., 2021 | 50,878 | Articles published in seventeen ecological journals between 1997 and 2017. | Ecology | Positive | On average research articles with more references are more cited than articles with less references and this difference is statistically significant. |
| | Hafeez et al., 2019 | 545 | Articles in 2007 from six major psychiatry journals in 2007. | Psychiatry | Positive | Number of cited references in articles significantly correlated with citation counts (rho=0.2) |
| | Haslam et al., 2008 | 308 | Articles published in top five economics journals between 2010 and 2014. | Psychology | Positive | Number and recency of cited references significantly correlated with citation counts (r=0.34 and 0.19, respectively). |
| | Peng & Zhu, 2012 | 7,749 | Articles from 105 journals in Internet studies. | Internet studies | Positive | Authoritativeness of cited references had significant positive correlation with citations ($\gamma$= 0.988, $p < 0.001$). |
| | Yu et al., 2014 | 1,025 | Articles published in 20 Library and Information Science journals. | Library & Information Science | Positive | The number of references significantly corelated with citation impact (rho=0.406 at the 0.01 level). |
| | Boyack & Klavans, 2005 | 780,049 | Articles from The Science Citation Index Expanded and The Social Sciences Citation Index during 2002 and 2003. | Multidisciplinary | Positive | Cited reference impact as measured by number of times the references from a certain paper were cited significantly correlated with citation counts, although there were disciplinary differences. |
| | Lancho-Barrantes et al., 2010 | 1.6 million | Articles from in 2007. | 27 Scopus subject categories | Positive | The averages of the SJR and JIF indicators are strongly correlated with the average number of references to recent papers included in the Scopus database. |
| | Bornmann et al., 2012 | 1,765 | Chemical articles. | Chemistry | Positive | h-index for the cited references significantly correlated with citation rates. |
| | Onodera & Yoshikane, 2015 | 1,395 | Sampled articles published in 2000 (n=230–240 in each field). | Five sciences and one engineering subject | Positive | The Price index (as measured by percentage of the references within five years before the publication year of the article) had the strongest association with the citation counts across fields ranging from rho=0.188 in electric and electronic engineering to rho=0.555 in biochemistry and molecular biology, followed by number of cited references ranging from rho=0.254 in condensed matter physics to rho= 0.494 in physiology. |
| | Lokker et al., 2008 | 757 | clinical articles from medical journals published from January to June 2005. | Medical Sci. | Positive | Citation counts of clinical articles were predicted by number of cited references [regression coefficient 0.004 (0.001 to 0.008)] in combinations with other factors. |

## 2. Predicting journal article citation counts from metadata

Many statistical studies have attempted to model factors that may associate with higher citation counts or have predicted long term citation counts from metadata rather than article text. For example, it is known that greater numbers of authors and country affiliations associate with more citations in many fields. Other studies have investigated the relationship between journal impact factors and article citation counts.

### 2.1. The relationship between the number of authors and citation counts

It seems reasonable to assume that articles with more authors tend to be better quality due to the greater range of expertise. Larger numbers of authors may also generate more interest for an article, an audience effect (Wagner, Whetsell, & Mukherjee., 2019). This section reviews research into whether different types of research collaboration tend to produce more highly cited articles. A positive association has been found in nearly all cases, but there is no agreed formula for the relationship between the two (e.g., linear, logarithmic).

**Articles with more authors are more cited (evidence form high impact journals):** A study of articles published in eight high impact multidisciplinary (Nature, Science and Proc. Natl. Acad. Sci.), biomedical (Circulation and Blood) and science (J. Am. Chem. Soc., Phys. Rev. Lett. and Astrophys. J.) journals during 1995 to 2004 found that co-authored papers tended to attract more citations. For instance, on average, a solo-authored Nature article had 61 citations, whereas Nature articles with 10 authors had 263 citations (Hsu & Huang, 2011). A similar report about the high impact journals Cell, Science, Nature, New England Journal of Medicine, The Lancet, and JAMA (n= 164 to 886) for the years 1975, 1985, and 1995 confirmed that the number of authors and the number of citations to articles positively correlated. For all journals, solo-authored research was the least cited (Figg et al., 2006).

**Articles with more authors are more cited (evidence from small studies):** Small studies on single fields have found that more authors associates with higher citation rates in Chemical Engineering (Peters & van Raan, 1994, n=226), Medical Sciences (Lokker et al., 2008, n=757), Psychology (Haslam et al., 2008, n=308), Pharmacology and Pharmacy (Bordons, Aparicio, & Costas, 2013, n=1,971 and 2,858), Ecology (Leimu & Koricheva, 2005, n=214), Library and Information Science (Sin, 2011, n=7,489), Computer Science (Ibanez, Bielza, & Larranaga, 2013, n=20,000), and management (Ronda-Pupo, 2017, n=36,241).

**Articles with more authors are more cited (evidence from large studies):** An early study of all papers from the Science Citation Index in Biomedical Research, Chemistry, and Mathematics in 1980, 1986, 1992, 1996, and 1998 found that multi-authored papers tended to attract more citations than solo papers (Glänzel, 2002). Another large-scale study of 19.9 million Web of Science articles in science and engineering (1955-2000), social sciences (1956-2000), and arts and humanities (1975-2000) found that articles with more co-authors received more citations than articles with individual authors across all broad fields and this citation advantage of co-authored research had increased over time. Although the majority (90%) of articles in the arts and humanities had single authors, co-authored papers tended to attract more citations and this difference was statistically significant (Wuchty, Jones, & Uzzi, 2007). A very large-scale study of 32.5 million Web of Science publications (articles, notes, and reviews) from 1900–2011 across the Natural & Medical Sciences and Social Sciences & Humanities found that more authors associated with higher citation impact in all fields. In Natural & Medical Sciences five authors and in Social Sciences & Humanities three authors on average were found to be important to start attracting substantially more citations (Larivière et al., 2015). Another large investigation across all 27 Scopus broad subjects from 10 countries with the most journal articles during 2008-2012 found that there was a significant increase in the average citation impact of research from single to two authored articles with a subsequent linear rise with additional authorship, giving overall logarithm-like shape (Thelwall & Maflahi, 2020). A paper about 226,166 Web of Science indexed articles in 2004 in Biology & Biochemistry, Chemistry, Mathematics and Physics found that the number of co-authors correlated with citation counts and the citation enhancement varied from 24% in Physics to 52% in Mathematics (Vieira & Gomes, 2010). A recent study of articles about Robotics and Artificial Intelligence (AI) (n=52,175) during 2008-2017 found that the number of authors and international collaboration have a significant low negative correlation with the waiting time to receive a first citation (r=-0.12 and -0.068 respectively), suggesting that as research collaboration increases, the waiting time for first citations decreases (Kumari et al., 2020).

**Formulae for the relationship between the number of authors and the expected citation counts of publications**: Although there is no agreement on the relationship between the number of authors of a paper and its expected citation impact, in general, the expected citation impact of a publication increases with the logarithm of the number of authors (based on all Scopus-indexed journal articles 2008-2012: Thelwall & Maflahi, 2020). The logarithm shape fits the UK well, for example (see Figure 10 of: Thelwall & Maflahi, 2020).

**Articles with more authors are more cited (evidence from a single country or institution)**: A study of Norwegian scientific production during 1981-1996 (n=46,849) in Natural Sciences found that highly cited articles tended to have more authors than normal papers. For instance, the average citation rate for articles with 10 authors was 4.5 times higher than for articles with solo articles (Aksnes, 2003). An investigation of Web of Science articles published in 2013 by Belgium (n=26,886), Israel (n=16,618) and Iran (n=28,203) found low but significant correlations between the number of authors and citation counts for most broad subjects. This included Chemistry (ranging from r=0.082 to r=0.105) and Clinical & Experimental Medicine (from r=0.161 to r=0.250), although there were differences between subjects and countries. For instance, in Social Sciences, while no significant correlation was found between co-authorship and citations for articles with Iranian addresses, the strongest association was found in this field for Israeli authors (r= 0.342) (Chi & Glänzel, 2017). At the country level, an investigation of 11,196 South African (Sooryamoorthy, 2009) and 15,301 Italian publications (Francescheta & Costantini, 2010) also found positive correlations between the number of authors and citations. Similarly, articles with more co-authors published by Harvard University during 2000-2009 (n=124,937) had more citations than single authored publications (Gazni & Didegah, 2011, $R^2$= 0.9). A large study across 27 broad subjects from the 10 countries with most journal articles during 2008-2012 found that increased collaboration associated with higher citation for all countries and subjects, except for China with a much lower association between academic collaboration rates and citation counts (Thelwall & Maflahi, 2020). Finally, a study of all Italian scientific production indexed in the Web of Science from 2004-2010 (n=392,257) also found significant associations between the number of authors and both citation impact and journal impact factors, with some disciplinary differences in the results (Abramo & D'Angelo, 2015).

**Articles with more authors are more cited (evidence from meta-analyses):** A recent meta-analysis of 92 relevant articles involving 340 effect sizes found a significant positive correlation, albeit weak, between research collaboration and citation counts (r=0.146), although this association was higher in Sciences & Biomedical and Social Science fields (both r= 0.167) than in other subjects. There was a stronger association between citations and research collaboration for developing countries (r= 0.180) than for developed countries (r= 0.112) (Shen et al., 2021).

**No evidence of associations between the number of authors and citation counts:** A few investigations have not found articles with more authors to be more cited. A study of 568 articles in eight economics journals in 1990 found no significant association between collaboration and citation counts (Medoff, 2003). Similarly, a study of 1,765 chemical articles papers in 2000 (Bornmann, Schier, Marx, & Daniel, 2012), an investigation of 2,792 articles from fourteen Finance journals during 1987–1991 (Avkiran, 1997) and an analysis of 50,162 articles in nanoscience and nanotechnology (2007-2009) found no significant correlations between the number of authors and citation counts (Didegah & Thelwall, 2013a). Similarly, no association was found in geography and forestry (n=213) (Slyder et al., 2011). Thus, in specific fields, co-authorship may not associate with more highly cited research. The same seems to be true for monographs. No association has been found between co-authorship of monographs (n=17,737) and their citation impacts, suggesting that collaboration indicators should not be used to predict scholarly impact in book-based fields (Thelwall & Sud, 2014).

## 2.2. The relationship between international collaboration and citations

Many studies have suggested that internationally co-authored papers tend to attract more citations compared with domestic articles. This may be due to wider audiences for the research (more people knowing the authors: Wagner, Whetsell, & Mukherjee, 2019), more varied expertise, or more funding (assuming that international collaboration is often triggered by grants). Conversely, a higher proportion of international research may be of highly funded types, whilst other international collaborations are more average. This section reviews relevant studies about this topic.

**Articles with international collaboration are cited more:** An early paper about astronomy papers (n=2,090, 1980-1991) found that on average articles with international collaboration tended to attract more citations than articles with national or no collaboration (Van Raan, 1998). For highly cited European physics articles from 1980-1987, about 41% had international collaboration (Glänzel et al. 1995). For 20,804 Web of Science articles in Sport Sciences during 2000–2001 and 2010–2011, articles with international co-authorship received more citations than articles with domestic co-authorship. Moreover, the relative citation impact of international publications was 1.16 and 1.29 for the periods 2000–2001 and 2010–2011, indicating that papers with international collaboration had higher citation scores than world average (1) in Sport Sciences (Wang,Thijs, & Glänzel, 2015). International collaboration also correlated with more citations for Web of Science articles between 2000-2009 in Biology and Biochemistry (n=16,058) and Chemistry (n=16,378) and one additional international collaboration associated with an increase the average citation count by 5.5% and 8.6% respectively (Didegah & Thelwall, 2013b).

**Articles with international authors are cited more (small-scale evidence):** Astronomy papers (n=2,090, 1980-1991) with international collaboration attracted more citations than articles with national or no collaborations (Van Raan, 1998).

**Articles with international authors are cited more (large-scale evidence):** An early study of 400,000 articles published between 1977 and 1986 in 28 subjects found that, on average, internationally co-authored papers (with more than one European Community country) attracted two times more citations than articles authored by a single country. The study also found that highly cited EU research tends to have multinational co-authorship (Narin et al. 1991). A large-scale study of 1.25 million articles from 1996-2012 in eight subject areas found that international collaboration had a significant positive association with citation counts, although varying between fields. For instance, in Ecology more than a quarter of articles with authors from five or more countries were within the 10% of most cited papers (Smith et al., 2014). Another large investigation of 32.5 million Web of Science publications in two broad research areas (Natural and Medical Sciences and Social Sciences and Humanities) also found that more international collaboration associates with increased citation impact for research both over time and between research areas (Larivière et al., 2015).

**Articles with international authors are cited more (evidence from one country or institution):** To assess if collaboration may associate with increased citation impact for UK academic research, an early study used a half million publications (between 1981 and 1991), finding that international and domestic collaborations associated with increases in citations by 1.6 and 0.75, respectively (Katz & Hicks, 1997). Similarly, a study of 46,849 Norwegian publications (1981-1996) in Natural Sciences found that about 63% of the highly cited papers were co-authored internationally compared with 26% overall (Aksnes, 2003). A report into 33,524 Scopus articles published in 2000 with at least two different author affiliations from European countries found that the geographical distance between the collaborating countries positively associates with citation counts. The regression analysis suggested that for each kilometre, citation counts increased by 7%-9% (Nomaler, Frenken, & Heimeriks, 2013). A study of 143,221 Finnish publications between 1990 and 2008 also showed a positive association between international co-authorship and citations and that articles produced via international collaboration tended to receive more citations than domestic co-authored research (Puuska, Muhonen, & Leino, 2014). A study of Web of Science articles from 2003-2013 found a negative association between citations and government funding for 35 OECD countries, but international collaboration had a significant and positive association with citation impact (Leydesdorff, Bornmann, & Wagner, 2019). International collaboration also positively associated with the overall citation impact of both young (n=26) and old universities (n=28) (Khor & Yu, 2016).

**Not all international collaboration is beneficial:** There is evidence that some countries may extract more value from international collaboration than others (Lancho-Barrantes, Guerrero-Bote, & de Moya-Anegón, 2013) and some countries may not benefit from international collaboration in terms of increased citation impact (Smith et al., 2014). For example, a study of articles published during 2004-2008 in the Nature and Science found that American authors publishing in these journals did not benefit from international collaboration in terms of citation impact (Rousseau & Ding, 2016). For Biochemistry articles in 2011 (n=13,578), research collaboration with the U.S. associated with increased scholarly impact for published research, whereas co-authorship with some other countries including India and China associated with reduced impact (Sud & Thelwall, 2016).

**No evidence of an association between international collaboration and citations:** International collaboration does not correlate with more citations for articles in the Social Sciences (n=15,932, 2000-2009) (Didegah & Thelwall, 2013b). A study of Harvard University publications 2000-2009 (n=124,937) also found no significant association between international collaboration and citation counts (Gazni & Didegah, 2010).

## 2.3. The relationship between institutional collaboration and citations

**Articles with more institutional collaborations are cited more:** An early study of UK publications (n= 376,000) between 1981 and 1991 found that domestic collaborations were associated with 0.75 more citations (Katz & Hicks, 1997). For 124,937 publications affiliated with Harvard University 2000-2009 there was a significant correlation between the number of collaborating institutions and citation counts (Gazni & Didegah, 2011, $R^2$= 0.72). For Pharmacology and Pharmacy articles by Spanish authors (n=1,971 for 1998-2000), articles with authors from different institutional sectors received more citations than articles with authors within the same institution (Bordons, Aparicio, & Costas, 2013, 2013). An investigation of 765,491 Web of Science articles in Artificial Intelligence 1997 to 2017 found that the type of institutional collaboration has a significant association with citations and there is significant association between citation counts and the number of “Main institutions type” (top 20 institutions in the field such as MIT, Stanford University, or University of Oxford), whereas no significant relationship was found between citations and the number of “Normal institutions type” (Fan et al., 2020). A report about nanoscience and nanotechnology journal articles from 2007-2009 (n=50,162) also found weak but significant associations between the number of collaborating institutions and citation counts (Didegah & Thelwall, 2013a). A very large investigation of over 32.5 million publications (1900–2011) in two research areas (Natural & Medical Sciences and Social Sciences & Humanities) also found that Web of Science publications with more author institutional addresses tended to have higher citation counts (Larivière et al., 2015). A study of six high impact multi-disciplinary and medical journals (Cell, Science, Nature, New England Journal of Medicine, The Lancet, and JAMA) for the years 1975, 1985, and 1995 found that the number of citations to articles correlated significantly with the number of institutions (Figg et al., 2006).

**No evidence of an association between institutional collaboration and citations:** A study of articles in Biology & Biochemistry (n=16,058), Chemistry (n=16,378) and Social Sciences (n=15,932) from 2000-2009 found no significant association between the number of collaborating institutions and the citation impact of the published research (Didegah & Thelwall, 2013b). In the field of Artificial Intelligence, no significant association was found between the citation impact of research and the number of “Normal institutions type” (Fan et al., 2020).

## 2.4. The relationship between journal impact factors and citations

Since the journal impact factor is calculated from the citation rates of the articles in a journal, it is logical to expect articles to be more cited when they are in a journal with a higher journal impact factor. This relationship is not certain, however, since individual highly cited articles may be the cause of a higher journal impact factor and the impact factor calculation only covers a limited range of citations. Nevertheless, there is strong evidence from many studies of different fields that there is a strong general relationship.

**Journal impact factors associate with higher citation rates for their articles (evidence from small studies):** An analysis of 204 articles in Emergency Medicine found that journal impact factors were the strongest predictor of citation counts ($R^2$=0.14) (Callaham, Wears, & Weber, 2002). For 196 articles published in five General & Internal Medicine journals with high impact factors in 2006, there was a significant medium Spearman correlation (rho= 0.63) between the impact factor of the journals and future article citations (Falagas et al. 2013). Similarly, in geography and forestry (n= 213) articles published in journals with higher impact factors also had more citations ($R^2$= 0.28) (Slyder et al., 2011) and an investigation of 131 articles in Environment and Ecology during 2006–2007 found that journal impact factors had a significant medium correlation (r=0.56) with citation counts (Vanclay, 2013). A study of 1,371 articles in demography (1990-1992) also found a significant and high association (r= 0.74) between journal impact factors and the number of citations to articles after 10 years (Van Dalen & Henkens, 2005). Using negative binomial regression models, a study of 1,586 articles in in biomedicine found the journal impact factor to be the most significant factor (coefficient= 0.11) to predict article citation counts (Bornmann & Daniel, 2006). For clinical systematic reviews and meta-analyses (n=1,261) published in 2008, journal impact factors could predict more than half ($R^2$=0.592) of the variation in their future citations (Royle et al., 2013).

**Journal impact factor associate with more citations (evidence from medium-large studies):** A study of 46,849 articles by Norwegian scientists 1981-1996 in Natural Sciences found that articles published in journals with high impact factors tended to be cited more. About 91% of the highly cited articles were published in journals with an impact factor above the field average (Aksnes, 2003). An investigation of immunology (n=13,125) and surgery (n=17,083) found significant high negative associations between the proportion of uncited articles and journal impact factors for both subject areas (rho= -0.854 and -0.924 respectively), suggesting that high impact journals published few uncited articles (Weale, Bailey, & Lear, 2004). Using machine learning models to predict citation counts, a study of 3,788 papers about internal medicine published between 1991 and 1994 found that the journal impact factor was the only content-based and bibliometric feature that ranked highly for all three studied citation thresholds, reporting absolute value of regression coefficients 4.04, 3.34 and 3.32 for citation thresholds 20, 50 and 100, respectively (Fu & Aliferis, 2010). For articles in Biology & Biochemistry (n=44,248), Chemistry (n=97,177), Mathematics (n=20,127) and Physics (n=64,614), the journal impact factor was the variable with the largest effect on citations (Vieira & Gomes, 2010). Internet studies articles (n=7,749) published in journals with higher impact factors also had more citations ($\gamma$= 0.537, $p < 0.001$) (Peng & Zhu, 2012). In nanoscience and nanotechnology the journal impact factor was the most significant factor associating with article citation counts (n=50,162), with a 1 SD increase in the impact factor associating with a 39% rise in citations to articles (Didegah & Thelwall, 2013a). A follow-up paper about articles in Biology and Biochemistry (n=16,058), Chemistry (n=16,378) and Social Sciences (n=15,932) published between 2000-2009 also found that the journal impact factor significantly correlated (rho= 0.455, 0.459, 0.186, respectively) with increased citations to articles in all three areas (Didegah & Thelwall, 2013b). Using the field-normalised average journal impact, a report about Pharmacology & Pharmacy (n=1,971 and n=2,858) also found that articles published in high impact factor journals were likely to attract more citations (Bordons, Aparicio, & Costas, 2013). A study of 9,898 papers from 2000 to 2004 matched with F1000 data found that journal impact factors had the strongest association with citations out of a range of different bibliometric and quality indicators (judgments of peers), discussing differences between qualitative judgments by experts and journal impact factor to predict future citations (Bornmann & Leydesdorff, 2015). A positive but weak correlation was also found between the journal impact factor and citation counts (r=0.327) in a recent paper about 9,823 articles published in 2016 and 2017 from 33 plastic surgery journals (Asaad et al., 2020). Finally, a large study of 780,049 Web of Science articles in 2002 and 2003 also showed that the journal impact factor had the strongest bibliometric correlation with citations (r=0.478) and a similar positive association was found in 17 out of 24 subject areas (Boyack & Klavans, 2005).

**No evidence of an association between journal impact factors and citation:** A few studies have found insufficient statistical evidence that articles published in journals with high impact factors tend to have higher citation impacts. These have covered Urology (Willis et al., 2011, n=200), Ecology (Leimu & Koricheva, 2005, n=214), and Gastroenterology and Hepatology (Roldan-Valadez & Rios, 2015).

## 2.5. Author publication and citation records and article citations

It seems reasonable to hypothesise that authors with a good track record of publishing or attracting citations would be more likely to author future highly cited papers. It is hard to fully assess this with career-level analyses, but there is some evidence in favour of the hypothesis.

**An author's h-index associates with citation counts:** A report about 1,025 articles in library and information science found low but significant correlations between the h-index of the first author and the maximum h-index of all authors with the citation impacts of their future articles (r=0.175 and 0.287, respectively) (Yu et al. 2014). A paper concerning 100,000 papers recommended by the China Computer Federation found that the maximum h-index of all authors associated with higher citation counts in computer science subjects (Qian et al., 2017). Similarly, a study of 219 articles in Astronomy & Astrophysics (Wang, Yu, & Yu, 2011) published in four journals in 1985 (Wang et al., 2012) and 1,860 papers written by 65 biomedical researchers (He, 2009) both found the h-index to be a significant predictor of citation counts. An investigation of 131 articles in Environment and Ecology (2006–2007) also found a medium correlation (r=0.42) between the maximum author h-index and citation counts. This association might be due to authors with higher h-indexes being likely to publish in high impact journals (Vanclay, 2013). For 18,000 publications by senior researchers from 147 chemistry research groups in the Netherlands during 1991-1998, there was a high significant correlation ($R^2$= 0.89) between the h-index and the total number of citations for all research groups. There were also associations between both the h-index and normalized citation impact and peer review judgment about the research quality of groups (van Raan, 2006). A large multidisciplinary Web of Science article study across 22 subjects for 2000-2009 found that a unit increase in the h-index associates with a 2.3% increase in citations for all studied subjects, although there were disciplinary differences, with the increase in citation counts being much higher in Mathematics (6.6%) and Economics & Business (5.1%) than in Immunology and Materials Science (both 0.8%) (Didegah, 2014).

## 2.6. Predicting the quality of journal articles based on machine learning

Several teams have used machine learning techniques to predict the long-term citation counts or quality scores of papers. An early attempt used Gradient Boosted Regression Trees to predict citation counts for 27,770 papers from arXiv high energy physics theory based on the contents, topics and author collaboration features of papers, finding that the models used were reasonably effective at predicting the future citation counts of papers (Chen & Zhang, 2015). Using machine learning on citation-based indicators (e.g., total citations and average h-index) and Times Higher Education indicators, an experiment assessed if REF 2014 overall university grade scores could be predicted. For this, 79 and 30 UK universities were divided into training and test sets respectively. The number of Web of Science publications, entry tariff and percentage of students were the most significant predictors university grade point averages (Balbuena, 2018), but the sample sizes used were very small for machine learning.

Most experiments in this section predicted future citation counts for articles from a field or set of journals. Machine learning models have been used to predict the future citations of biomedical research (1991-1994) in six medical journals (JAMA, Lancet, NEJM, BMJ, American Journal of Medicine, and Annals of Internal Medicine). Overall, 3,788 documents, 20,005 article text features (article title, abstract, MeSH terms, publication type), metadata (number of authors and institutions, number articles for first and last authors in the previous 10 years, quality of first author's institution) and citations (number of citations for first and last authors, Journal impact factor) were leveraged. First author citations had the greatest association (coefficient=5.75) with articles reaching a citation threshold 100, followed by the MeSH topic Smoking: mortality (4.22), journal impact factor (3.32) and last author citations (3.02). Overall, the study suggested that it is feasible to predict future citation counts with machine learning techniques to some extent (Fu & Aliferis, 2010). Other small-scale studies have used machine learning and different article or metadata features to predict citation counts, such as from machine learning conference papers (Li et al., 2019; Cummings & Nassar, 2020), articles from the selected journals (e.g., Wang et al., 2020; Zhao & Feng, 2022) or papers on a specific topic (Xu et al., 2019).

Long term citation counts can be predicted from early citation counts and/or metadata. Deep learning techniques have been used on a dataset of articles published in Nature, Science, The New England Journal of Medicine, Cell and Proceedings of the National Academy of Sciences (n=175,432) to predict long-term counts of articles based on citation counts soon after publication (Abrishami & Aliakbary, 2019). A similar study used deep learning to predict the 5-year citation impact of library, information and documentation articles (n=49,834) from Chinese Social Sciences Citation Index (2000 to 2013). The study applied multiple features from article text (document type, article length, title length, funding, month of publication, punctuation in the title), journal (journal impact factor and number of publications in the journal), authors (e.g., number of authors, productivity, previous citations, h-index and number of organizations), references (e.g., number and age of references, self-citations and percentage of different document types in references) and citations (citations in the first or first two years, number of citing journals in the first or first two years), with some positive results (Ruan et al., 2020).

One study used metadata semantic features from Artificial Intelligence (AI) related articles published in 20 journals indexed by China Computer Federation catalogue to predict the future citation impact of papers with deep learning techniques for semantic features extraction in the AI subject (Ma et al., 2021). In contrast to above studies, a study used multiple altmetric indicators (e.g., Mendeley reader, open peer-review shares, or mentions in Twitter, news or blogs) in addition to other metadata to predict future citations for a random sample of 12,374 articles published in 2015. Using machine learning models, the study found that Mendeley readership, maximum followers on Twitter, and academic status (e.g., student, postdoc, researcher, or professor) were top parameters to predict the short-term and long-term citations impact of papers (Akella et al., 2021). All of the papers in this section used relevant inputs to predict future or long-term citations but because of the different datasets used, it is not possible to generate general conclusions about which methods or inputs are best overall or in particular cases.

A very large multi-disciplinary study used 32 different machine learning methods and all Scopus journals published during 2014 to 2020 across 326 Scopus narrow subjects to predict the quality of published research. Citations (the Normalized Log-transformed Citation Score), collaboration (Number of authors and number of country affiliations) and article text (words from the title, abstract, and keywords) were used as inputs for the machine learning process. The study found that two machine learning methods (Gradient Boosting Classifier and Random Forest Classifier) had the highest levels of accuracy compared with other methods (46% and 45% respectively) and machine learning can predict the citation-based journal third of articles using the selected features (Thelwall, 2022b).

Machine learning methods have also been used to identify the characteristics of highly cited articles. A study extracted features from article text (title length, number of figures, tables, equations, and characters with no spaces), metadata (number of authors and number of views) and citation counts from high and low cited papers (each 100 articles, n=200) published by MDPI in 2017, finding significant positive associations between citation counts and the number of views, tables, and authors and a negative but significant correlation with title length (Elgendi, 2019).

A study of the association between REF 2014 grade point averages and impact factors (IF) in Neuroscience, psychiatry, and psychology found that the proportions of publications ranked 4* and 3* can be predicted with 95% and 98% accuracy (Al-Janabi, Lim, & Aquili, 2021).

# Table 2. Summary of studies predicting journal article citation counts from metadata

| Factors | Study | Number of articles | Dataset | Subject | Association with citations | Main result |
|---|---|---|---|---|---|---|
| Number of authors and citation impact | Hsu & Huang, 2011 | 92,034 | Articles published in eight high impact multidisciplinary, biomedical and science journals during 1995 to 2004. | Multidisc. Biomed. Science | Positive | Nature solo articles had 61 citations, whereas articles with 10 (or more) had 263 (or 370) citations. |
| | Figg et al., 2006 | 8,631 | Articles from six high impact multidisciplinary and biomedical journals for the years 1975, 1985, and 1995. | Multidisc. Biomed. | Positive | Articles with more authors had more citations and solo-authored papers had the least citation counts. |
| | Glänzel, 2002 | All papers from SCI | Papers from Science Citation Index in Biomedical Research, Chemistry, and Mathematics in the years 1980,1986,1992, 1996, and 1998. | Biomed. Chem. Math. | Positive | Multi-authored papers tended to attract more citation than solo author papers in the three selected fields. |
| | Wuchty et al., 2007 | 19.9 million | Web of Science articles in science and engineering, social sciences and arts and humanities. | Science Social Sci. Humanities | Positive | Articles with more co-authors had more citations than articles with solo authors across all broad fields. |
| | Larivière et al., 2015 | 32.5 million | Web of Science publications (1900–2011) across two broad research areas (Natural & Medical Sciences and Social Sciences & Humanities). | Natural & Medical Sci. Social Sci. & Humanities | Positive | Author counts associated with citation counts in all research fields. |
| | Thelwall & Maflahi, 2020 | - | Ten countries with most journal articles during 2008-2012 in 27 broad subjects. | Multidisc. | Positive | A linear association between co-authorship and citation impact. |
| | Vieira & Gomes, 2010 | 226,166 | Web of science articles in 2004 in Biology & Biochemistry, Chemistry, Mathematics and Physics. | Biochem. Chem. Math. Phys. | Positive | The number of co-authors correlated with citation counts and the citation enhancement varied from 24% in Physics to 52% in Mathematics. |
| | Kumari et al., 2020 | 52,175 | Articles about Robotics and Artificial Intelligence (AI) related subjects during 2008-2017. | Robotics and AI | Positive | Articles with more authors and international collaboration attract citations faster than other articles. |
| | Aksnes, 2003 | 46,849 | Articles by Norwegian authors 1981-1996. | Natural Sciences | Positive | The average citation rate for articles with 10 authors was 4.5 higher than for solo articles. |
| | Chi & Glänzel, 2017 | Belgium (26,886) Israel (16,618) Iran (28,203) | Web of Science articles published in 2013 by Belgium, Israel and Iran. | Multidisc. | Positive | Significant but low correlations between the number of authors and citations for most broad subjects and three countries such as in Chemistry (ranging from r=0.082 to r=0.105) and Clinical & Experimental Medicine (from r=0.161 to r=0.250). |
| | Shen et al., 2021 | Meta-analysis | Meta-analysis of 92 relevant articles involving 340 effect sizes. | Multidisc. | Positive | A significant but weak association between collaboration and citation counts (r=0.146), higher in Sciences & Biomedical and Social Science fields (both r= 0.167) and for developing countries (r= 0.180) than developed countries (r= 0.112). |
| | Medoff, 2003 | 568 | Articles published in eight economics journals in 1990. | Economics | No association | No significant association between collaboration and citation counts. |
| | Bornmann et al., 2012 | 1,765 | Article in Chemistry published in 2000. | Chemistry | No association | No significant correlation between co-authorship and citation counts. |
| | Avkiran, 1997 | 2,792 | Articles from fourteen Finance journals during 1987–1991. | Finance | No association | No significant correlation between the number of authors and citation counts. |
| | Didegah & Thelwall, 2013a | 50,162 | Articles in nanoscience and nanotechnology in 2007-2009. | Nanosci. & nanotech. | No association | No significant association between the number of authors and citation counts. |
| | Slyder et al., 2011 | 213 | Articles in geography & forestry up to the year 2010 from ten American universities. | Geography & forestry | No association | No relationship between the number of authors and citation counts. |
| International collaboration and citation impact | Narin et al. 1991 | 400,000 | Articles published between 1977 and 1986 in 28 subjects. | Multidisc. | Positive | Internationally co-authored papers had two times more citations than articles authored by a single country. |
| | Smith et al., 2014 | 1.25 million | Articles between 1996-2012 in eight subject areas. | Chem., Phys., Ecol., Bio., Gene. Geol., Math., Psych. | Positive | Articles with authors from more countries were cited more. In Ecology more than a quarter of articles with authors from five or more countries were within 10% of mostly cited papers. |
| | Larivière et al., 2015 | 32.5 million | Web of Science publications (1900–2011) across two broad research areas (Natural & Medical Sciences and Social Sciences & Humanities). | Natural & Medical Sci. Social Sci. & Humanities | Positive | More international collaboration can subsequently increase citation impact of research both over time and between research areas. |
| | Van Raan, 1998 | 2,090 | Articles in Astronomy published during 1980-1991 1977 in 28 subjects. | Astronomy | Positive | International collaboration attracted on more citations than articles with national or no collaborations. |
| | Wang et al., 2015 | 20,804 | Web of Science articles in Sport Sciences during 2000–2001 and 2010–2011. | Sport Sciences | Positive | Relative citation impact of international publications was 1.16 and 1.29 compared to domestic co-authorship for the periods 2000–2001 and 2010–201 respectively. |
| | Didegah & Thelwall, 2013b | 16,058 (Bio. Sci.) 16,378 (Chem.) | Web of Science indexed articles during 2000-2009. | Biology & Biochem. Chemistry | Positive in Biology Chemistry. | One additional found to increase the average citation count by 5.5% and 8.6% in Biology and Biochemistry and Chemistry respectively. |
| | Katz & Hicks, 1997 | 376,226 | UK publications between 1981 and 1991. | General | Positive | International collaborations increase citations by 1.6. |
| | Aksnes, 2003 | 46,849 | Articles by Norwegian authors 1981-1996. | Natural Sciences | Positive | About 63% of the highly cited papers were co-authored internationally compared 26% overall . |
| | Nomaler et al., 2013 | 33,524 | Scopus articles published in 2000 from the European countries. | General | Positive | Geographical distance between collaborating countries positively associates with citation counts. |
| | Puuska et al., 2014 | 143,221 | Finnish publications between 1990 and 2008 in six broad fields (natural science, medicine, engineering, agriculture, social sciences humanities). | Six broad fields. | Positive | Iinternational collaborations tended to receive more citations than domestic co-authored research. |
| | Leydesdorff et al., 2019 | - | Web of Science articles during 2003-2013. | General | Positive | International collaboration had a significant and positive association with citation impact. |
| | Sud & Thelwall, 2016 | 13,578 | Biochemistry articles in 2011. | Biochemistry | Positive | Research collaboration with the U.S. increase citation impact, whereas co-authorship with some other countries may reduce it. |
| | Didegah & Thelwall, 2013b | 15,932(Soc. Sci.) | Web of Science indexed articles during 2000-2009. | Social Sci. | No association | No meaningful association between international collaboration and citation rates in Social Sciences. |
| | Gazni & Didegah, 2010 | 124,937 | Harvard University publications between 2000-2009. | General | No association | No significant relationship between international collaboration and citation counts. |
| Institutional collaboration and citation impact | Larivière et al., 2015 | 32.5 million | Web of Science publications (1900–2011) across two broad research areas (Natural & Medical Sciences and Social Sciences & Humanities). | Natural & Medical Sci. Social Sci. & Humanities | Positive | Web of Science publications with more institutional addresses of authors tended to have higher citation impact over time and in the studied research areas. |
| | Fan et al., 2020 | 765,491 | Web of Science articles in Artificial Intelligence subject between 1997 to 2017. | Artificial Intelligence | Positive | Significant association between citation counts and number of "Main institutions type" (top 20 institutions in the field). |
| | Gazni & Didegah, 2010 | 124,937 | Harvard University publications between 2000-2009. | General | Positive | Significant correlation between the number of collaborating institutions and citation counts. |
| | Bordons et al., 2013 | 1,971 | articles by Spanish researchers in Pharmacology and Pharmacy during 1998-2000. | Pharmacology and Pharmacy | Positive | Authors from different institutional sectors received more citations than articles with authors within the same institution. |
| | Didegah & Thelwall, 2013a | 50,162 | Articles in nanoscience and nanotechnology in 2007-2009. | Nanosci. & nanotech. | Positive | Weak association between the number of collaborating institutions and citation counts. |
| | Gazni & Didegah, 2010 | 124,937 | Harvard University publications between 2000-2009. | General | Positive | Significant correlation between the number of collaborating institutions in the published research and their citation counts. |
| | Figg et al., 2006 | 8,631 | Articles from six high impact multidisciplinary and biomedical journals for the years 1975, 1985, and 1995. | Multidisc. Biomed. | Positive | Articles with more authors had more citations and solo-authored papers had the least citations. |
| | Katz & Hicks, 1997 | 376,226 | UK publications between 1981 and 1991 | General | Positive | Domestic collaborations increase citations by 0.75. |
| | Didegah & Thelwall, 2013b | 15,932(Soc. Sci.) | Web of Science indexed articles during 2000-2009. | Social Sci. | No association | No significant association between the number of collaborating institutions and citation impact of published research. |
| | Fan et al., 2020 | 765,491 | Web of Science articles in Artificial Intelligence subject (1997 to 2017) | Artificial Intelligence | No association | No significant relationship found between citations and the number of "Normal institutions type". |
| Journal Impact factor and citation counts | Aksnes, 2003 | 46,849 | Articles by Norwegian authors 1981-1996. | Natural Sciences | Positive | Articles in journals with high impact factors tended to be cited more. About 91% of the highly cited articles were published in journals with an impact factor above the field average. |
| | Weale et al., 2004 | 13,125 (Immun.)<br>17,083 (Surgery) | Articles in immunology and Surgery. | Immunology and Surgery | Negative (between JIF and zero citation) | High negative associations between the proportion of uncited articles and journal impact factors for both immunology (rho= -0.854) and surgery (rho= -0.924), suggesting that high impact journals published few uncited articles. |
| | Fu & Aliferis, 2010 | 3,788 | Articles about internal medicine published between 1991 and 1994. | Medical Sci. | Positive | Journal impact factor was the only feature that ranked highly for all three studied citation thresholds (20, 50 and 100), reporting absolute value of regression coefficients 4.04, 3.34 and 3.32 respectively. |
| | Peng & Zhu, 2012 | 7,749 | Articles from 105 journals in Internet studies. | Internet studies | Positive | Higher impact factor journal articles had more citations ($\gamma$= 0.537, $p <$ 0.001). |
| | Vieira & Gomes, 2010 | 44,248 (Bio.)<br>97,177 (Chem.)<br>20,127 (Maths)<br>64,614 (Physics) | Web of Science articles in four science fields in 2004. | Bio. Biochem.<br>Chemistry<br>Maths<br>Physics | Positive | The journal impact factor was the variable with the largest effect on citation counts across all fields. |
| | Didegah & Thelwall, 2013a | 50,162 | Articles in nanoscience and nanotechnology in 2007-2009. | Nanosci. & nanotech. | Positive | The journal impact factor was the most significant contributing factor of citation counts of articles. A 1 SD increase in the impact factor relates to a 39% rise in citations to articles. |
| | Didegah & Thelwall, 2013b | 16,058 (Bio. Sci.)<br>15,932 (Soc. Sci.)<br>16,378 (Chem.) | Web of Science indexed articles during 2000-2009. | Biology & Biochem.<br>Social Sci.<br>Chemistry | Positive | The journal impact factor significantly correlated with increased citations of articles in all three subject areas (rho= 0.455, 0.459, 0.186, respectively). |
| | Bordons et al., 2013 | n=1,971 and 2,858 | Pharmacology and Pharmacy Web of Science articles by Spanish authors during 1998-2000 and 2006-2008. | Pharmacology and Pharmacy | Positive | Articles in high impact factor journals are likely to attract more citations. |
| | Bornmann & Leydesdorff, 2015 | 9,898 | Papers from 2000 to 2004 matched with F1000 data. | | Positive | Journal impact factor had the strongest association with citations among other studied bibliometric and quality indicators (judgments of peers). |
| | Asaad et al., 2020 | 9,823 | Articles published in 2016 and 2017 from 33 plastic surgery journals. | Plastic surgery | Positive | A positive moderate correlation between the journal impact factor and citation counts (r= 0.327). |
| | Boyack & Klavans, 2005 | 780,049 | Articles from The Science Citation Index Expanded and The Social Sciences Citation Index during 2002 and 2003. | Multidisciplinary | Positive | The journal impact factor has the strongest correlation with citations (r= 0.478) and a similar positive association found in 17 out of 24 subject areas. |
| | Van Dalen & Henkens, 2005 | 1,371 | Articles in demography published during 1990-1992. | Demography | Positive | A significant and high association (coefficient: 0.74) between journal impact factors and citations to articles after 10 years. |
| | Bornmann & Daniel, 2006 | 1,586 | Articles in biomedicine published by postdoctoral researchers. | Biomedicine | Positive | Journal impact factor was the most significant factor (Coefficient= 0.11) to predict citation counts of articles. |
| | Royle et al., 2013 | 1,261 | Clinical systematic reviews or meta-analysis published in 2008. | Med. Sci. | Positive | Journal impact factor could predict about more than half ($R^2$=0.592) of the variation in citations. |
| | Callaham et al., 2002 | 204 | Articles in Emergency Medicine | Emergency Medicine | Positive | Journal impact factor was the strongest predictor of citations ($R^2$=0.14). |
| | Falagas et al. 2013 | 196 | Articles published in five General and Internal Medicine journals with highest impact factor in 2006. | Med. Sci. | Positive | Significant and medium correlation (rho= 0.63) between the impact factor of journals and future article citations. |
| | Slyder et al., 2011 | 213 | Articles in geography & forestry up to the year 2010 from ten American universities. | Geography & forestry | Positive | Articles published in journal with higher impact factors also had more citations ($R^2$= 0.28). |
| | Vanclay, 2013 | 131 | Articles in Environment and ecology during 2006–2007. | Environment and ecology | Positive | Journal impact factor had the strongest correlation (r=0.56) with citation counts. |
| | Willis et al., 2011 | 200 | Articles published between January and June 2004 from three Urology journals. | Urology | No association | No association between articles published in journals with high impact factors and citations. |
| | Leimu & Koricheva, 2005 | 214 | Web of Science indexed articles in Ecology. | Ecology | No association | No relationship between journal impact factor and citation counts. |
| | Roldan-Valadez & Rios, 2015 | - | Web of Science indexed articles from 74 Gastroenterology and Hepatology journals. | Gastroent. & Hepatol. | No association | No association between journal impact factor and citation impact. |
| h-index and citation impact | Didegah, 2014 | Large study across 22 fields | Multidisciplinary study of Web of Science articles across 22 subjects during 2000-2009. | Multidiscip. | Positive | A unit increase in the h-index predicts a 2.3% increase in article citations for all studied subjects. |
| | Qian et al. 2017 | 100,000 | Recommended papers by the China Computer Federation. | Computer Science | Positive | The maximum h-index of all authors associated with the higher citation counts in computer science subjects. |
| | van Raan, 2006 | 18,000 | Papers by senior researchers from 147 chemistry research groups in the Netherlands during 1991-1998. | Chemistry | Positive | High significant correlation ($R^2$= 0.89) between the h-index and the total number of citations for all research groups. |
| | He, 2009 | 1,860 | Papers written by 65 biomedical researchers. | biomedical | Positive | h-index is a significant predictor for citation counts. |
| | Yu et al., 2014 | 1,025 | Articles published in 20 Library and Information Science journals. | Library & Information Science | Positive | Significant correlation between the h-index of the first author and the maximum h-index of the authors with citation impact of future articles (r=0.175 and 0.287, respectively). |
| | Vanclay, 2013 | 131 | Articles in Environment and ecology during 2006–2007. | Environment and ecology | Positive | Significant correlation (r=0.42) between the maximum author h-index and citation counts. |
| | Wang et al., 2011 | 219 | Articles in astronomy and astrophysics. | Astronomy | Positive | h-index was significant predictor for citation counts. |

# 3. Predicting or assessing journal article quality scores from metadata

There are few empirical studies about the estimation or prediction of the quality (however defined) of individual academic publications using automatic or semi-automatic methods, such as based on metrics. This is due to the lack of published large-scale expert judgements on articles or other research outputs. Nevertheless, some studies have used alternative approaches, such as predicting quality profiles or averages from published summary or aggregate data (e.g., institution-UoA departmental quality profiles for previous REFs and RAE 2008 or departmental numerical/star ratings for RAE 1992/1996/2001).

## 3.1. UK RAE/REF scores and bibliometric indicators

Many investigations of the relationship between departmental citation-based and journal-based bibliometric indicators and departmental average RAE/REF scores or score profiles have found statistically significant positive correlations. None have found a method that is capable of closely predicting the average scores (rather than rankings), however. For REF correlations, it is important to distinguish between those based on total output scores and average output scores. The latter will usually be lower, given that larger institutions tend to get higher scores in the UK.

**RAE 1992 scores associate with citation counts in Library and Information Science, Anatomy, Genetics and Archaeology:** Several studies investigated associations between Research Assessment Exercise (RAE) scores and citation metrics. An early study of RAE 1992 outputs from UK Library & Information Science (LIS) academics (n=217) found a significant and strong Spearman correlation between both the numbers of citations received by library and information science departments and the numbers of citations per member of staff and departmental RAE ratings (rho=0.81 and 0.82, respectively, significant at the p=0.01 level). The author concluded that “the cost and effort of the Research Assessment Exercise may not be justified when a simpler and cheaper alternative, namely a citation counting exercise, could be undertaken” (Oppenheim, 1995, p. 18). High significant correlations were also found between RAE 1992 ratings of Library and Information Science departments and total citations, average citations per staff member, and average citations per publication (Seng & Willett, 1995). A paper about multiple disciplines, including Anatomy, Genetics and Archaeology, also found high statistically significant Spearman correlations between 1992 RAE rankings and the total number of citations by departments (rho=0.718, 0.794 and 0.823, respectively) as well as medium-high correlations between RAE ratings and average number of citations per academic of staff (rho=0.487, 0.680 and 0.740 respectively) (Oppenheim, 1997). In the field of Business & Management Studies, another study found a significant correlation (r=0.682) between journal ranking scores (Discipline Contribution Scoring) and the 1992 RAE rating (Thomas & Watkins, 1998).

**The RAE 1996 score associates with average citations in Psychology:** One unpublished report about the field of Psychology (n academics=747) found a very high Spearman correlation (rho=0.91) between the 1996 RAE ratings and mean departmental citations (Smith & Eysenck, 2002).

**The RAE 2001 score associates with average citation counts in Archaeology, Music, Psychology and Political Science:** A paper about Archaeology (n academics=692) found high significant correlations between the 2001 RAE ranking score and total staff average citations (rho=0.85) and the total of all staff citations (rho=0.79) (Norris & Oppenheim, 2003), which was almost the same as result in the

same subject area for 1992 RAE (0.740 and 0.823 respectively, see also above for Oppenheim, 1997). The 2001 RAE scores for UK university Music departments highly correlated with departmental total and average citation counts (rho=0.80 and 0.81 respectively), although a weaker correlation was found between RAE scores and individual citation counts (rho=0.46) (Oppenheim & Summers, 2008). In Psychology, a strong association (rho=0.86) was also reported between the 2001 RAE scores and mean departmental citations (Smith & Eysenck, 2002). A regression study of 4,400 submissions to the 2001 RAE Political Science panel found that the mean number of citations to the submitted works were the most significant predictor of the RAE scores for the 69 political science departments (Partial and standardised coefficients: 0.541 and 0.340 respectively). This study also investigated which document types were more likely to attract citations, finding that journal articles were the most significant publication type in predicting the RAE outcome (Partial and standardised coefficients: 0.585 and 0.427 respectively) compared with authored books (0.151 and 0.223) or book chapters (0.115 and 0.250), although the 2001 RAE advice was that authored books would be rated higher than journal articles in the Political Science panel (Butler & McAllister, 2009).

**Mixed results for associations between the RAE 2001 scores and citation metrics:** A large and multidisciplinary study of RAE 2001 research outputs (n= 112,201 or 55% of all submissions) found positive and statistically significant correlations between departmental Web of Science citations and departmental RAE peer review score profiles across most science subjects, but no significant association was found in most social science and humanities subjects. The Spearman correlations between departmental average numbers of citations and RAE 2001 scores were high and significant in most biomedical fields (7 out of 8) ranging from 0.821 in Clinical Laboratory Sciences to 0.573 in Hospital-based Clinical Subjects, except for Nursing, where insignificant associations were found. However, in Physical and Engineering subjects, there were more diverse relationships between citation and RAE scores (significant in 9 out of 13 subjects) and associations were higher in Chemistry, Earth Sciences, and Physics (0.789, 0.754 and 0.685, respectively), whereas no significant associations were found for Pure Mathematics, Civil Engineering, Electrical & Electronic Engineering and Mechanical Engineering. The correlations were not significant in most Social Science subjects (e.g., Politics, Sociology and History), except for Business & Management (0.782), Economics & Econometrics (0.677) and Geography (0.383) (Mahdi, D'Este & Neely, 2008). This study suggested that there were some disciplinary differences in the relationships between departmental average peer judgment scores and citations.

**RAE 2008 rankings associate with citations in Chemistry and Political Science:** Butler and McAllister (2011) used a method that had been previously applied to the RAE 2001 Political Science category (see above Butler & McAllister, 2009), but this time to predict the RAE 2008 outcome in Chemistry (n=44 departments) and Political Science (n=69 departments). They again found that citation metrics could be useful indicators for predicting RAE outcomes in Chemistry (Partial and standardised coefficients: 0.491 and 0.370 respectively) and in Political Science (0.410 and 0.273) (Butler & McAllister, 2011).

**The RAE 2008 rankings associate with Article Influence Score in Sociology:** A study used three variables (quality of journals in the submission, research income per capita and scale of research activity), predicting about 83% of the variance in RAE 2008 outcomes for Sociology. The impact of journals was assessed using an indicator called Article Influence Score[1] (n=2,366), finding significant associations (beta coefficients of 0.37) between the journal quality indicator and the RAE 2008 average scores for sociology (Kelly & Burrows, 2011).

**Weak evidence of an association between the RAE 2008 scores and citations:** A multi-disciplinary study found varied Spearman correlations between RAE 2008 scores for research groups and citation counts across the selected subjects, with weaker associations found for Mechanical, Aeronautical & Manufacturing Engineering (rho= 0.18), History (0.38), Sociology and Geography & Environmental Studies (both 0.47) than Physics and Biology (both 0.57) and Chemistry (0.62). Because of the relatively low-medium correlations between citations and average RAE scores, the study argued that citation-based metrics are "a poor proxy for peer-reviewed measures of the quality of research groups" (Mryglod et al., 2013, p.10).

**RAE 2008 outcomes associate with journal quality scores in Business, Management and Economics:** A report about the RAE 2008 outputs found high significant correlations between departmental RAE ratings and the UK's Association of Business Schools (ABS) Journal Quality Scores in Business & Management (0.773) and Economics & Econometrics (0.704) at 0.1% level, arguing that "Requiring the panels to take bibliometric indicators such as journal quality scores into account should help not only to reduce their workload but also to mitigate the implicit bias indicated by the statistical analyses reported in this paper" (Taylor, 2011).

**RAE 2008 scores associate with the reputations of journals and book publishers in Political Science:** The reputations of political science journals and book publishers (as measured by a survey of British political scientists) associated with the departmental proportions of top-rated scholarly outputs in the 2008 RAE. For instance, submitted outputs in top 10 journals based on reputational surveys were moderately correlated with the proportions of 4* (rho=0.49) and 3* (0.33) ratings, whereas this was negative for 2* and 1* rated research (-0.15 and -0.43 respectively). The proportions of non-top 20 journals in Political Sciences had significant negative correlations with the proportions of 4* (-0.48) and 3* (-0.35) RAE ratings. Similar associations were found between the proportions of articles in the top 20 journals and RAE ratings. The departmental proportion of monographs from top publishers also associated with higher proportions of 4* (0.78) and 3* (0.42) ratings and lower proportions of 2* (-0.37) and 1* (-0.58) RAE ratings (Allen & Heath, 2013).

**RAE 2008 scores associate with Google Books citations to books in Communication Cultural, and Media studies:** To investigate whether citations from the huge Google Books database might be helpful for research assessment, a study found a weak, but significant Spearman correlation (rho=0.387) between the 2008 RAE average ranking scores in Communication, Cultural, and Media Studies for 47 institutions and average Google Books citations to the 407 books that they had submitted. Since books tend to be much longer than journal articles, even weak evidence from Google Books citation counts might be helpful to support the peer-review process (Kousha, Thelwall, & Rezaie, 2011).

[1] https://jcr.help.clarivate.com/Content/glossary-article-influence-score.htm#:~:text=The%20Article%20Influence%20Score%20determines,all%20articles%20in%20all%20publications

**RAE/REF scores associate with departmental h-indexes:** High significant correlations have been found between departmental RAE 2008 grade point averages and departmental h and g index scores in Pharmacy (0.772 and 0.696 respectively). The association was weaker in Library & Information Management (0.397 and 0.378) and in Anthropology this association was negative (Norris & Oppenheim, 2010). For REF 2014, stronger Pearson and Spearman correlations were found between departmental h-indexes and different REF score weightings in Biology (ranging from 0.71 to 0.79), Chemistry (0.71 to 0.83), Physics (0.44 to 0.59) and Sociology (0.53 to 0.62) than with institutional normalized citation impact (ranging from 0.37 to 0.67 in different fields). This study argued that the h-index could be better citation indicator to predict REF outcome (Mryglod et al., 2015). A blog post also argued that departmental h-indexes could predict RAE 2014 results in Psychology (Bishop, 2014).

**Mixed results for associations between the RAE 2014 scores and citations:** A study by Elsevier found a moderate correlation (0.59) between universities' proportions of 4* outputs (world-leading) in the 2014 REF and the proportion of their articles that were in the global top 5% highly cited. However, there were large disciplinary differences, with the association being much higher in Biological Sciences, Chemistry, Psychology, Psychiatry & Neuroscience, Business & Management Studies and Computer Science & Informatics (r≈0.7 to 0.75) than in other fields, and the association was very weak in Physics and Clinical Medicine (up to r≈0.3) (Jump, 2015).

**Large-scale individual publication-level assessment of the REF 2014 scores and bibliometric indicators:** Unlike all previous investigations of associations between the RAE/REF peer review scores and citation metrics at the departmental level, a large-scale study of the 78% (n=149,670) of the research outputs with DOIs from REF 2014 assessed correlations with REF 2014 peer review scores at the article level, although reporting results from the oldest REF 2014 year, which was 2008, in the greatest detail. For the first time, 10 bibliometric and 5 altmetric indicators across all 36 REF 2014 subjects were used to predict the REF outcomes. Overall, the results showed that REF peer review scores for individual articles significantly and positively correlated with most of the selected bibliometric indicators, including with SCImago Journal Rank (rho=0.340), Source-normalised impact per paper-SNIP (0.327), field-weighted citation impact (0.284), and number of citations per publication (0.246). However, the study showed that the publication year of the research submitted to REF could significantly influence the results. For instance, Spearman correlations between REF peer review scores with number of citations per publication and were much higher for 2008 publications than for 2013 (0.382 and 0.154 respectively) because older submitted research to REF had more time to be read and cited. Moreover, the study reported huge disciplinary differences in the associations between REF scores and bibliometric indicators. For instance, the number of citations per publication for the 2008 dataset had the strongest associations with REF scores (above 0.5) in Clinical Medicine (rho=0.676), Chemistry (0.609), Physics (0.608) and Biological Sciences (0.589), whereas in almost all the social sciences and the art and humanities subjects there were very low (rho under 0.3) or statistically insignificant correlations between variables. In Economics & Econometrics, the SCImago Journal Rank (0.751) and source normalised impact per paper (0.665) had the strongest associations with REF scores. The study concluded that in most medical and science fields in main panels A and B, many bibliometric indicators associated with REF peer review scores, although they could not predict REF peer review with sufficient precision and sensitivity (HEFCE, 2015). In the executive summary relevant to the above study, it was recommended that "peer review, despite its flaws and limitations, continues to command widespread support across disciplines. Metrics should support, not supplant, expert judgement" (Wilsdon et al., 2015a).

**Publication-level vs. institutional level assessment of correlations between RAE/REF scores and bibliometric indicators:** Using a similar methodology to an earlier investigation (Mahdi, D'Este & Neely, 2008), two-thirds of 2014 REF outputs were matched with Web of Science records (133,469 out of 190,962) and different measures were used to assess the agreement between metric-based departmental rankings and REF peer review departmental rankings. There were very high Pearson correlations (r higher than 0.8) between the percentages of 4* rated submissions and the percentage of top 10% publications in Economics & Econometrics, Clinical Medicine, Physics, Chemistry, and Public Health. This association was also relatively high (at least 0.7) in Earth Systems & Environmental Sciences, Psychology, Psychiatry & Neuroscience, and Electrical & Electronic Engineering, Metallurgy & Materials. Overall, the associations between citation metrics and REF scores were higher at the departmental level than at the publication level as reported by the HEFCE study (see above HEFCE, 2015), presumably due to averaging effects. Another investigation suggested that top percentile of most cited papers from the UK universities may substitute for REF peer review in Chemistry, Economics & Econometrics, Business & Management Studies, and Physics (Rodríguez-Navarro & Brito, 2020).

**REF 2014 scores associate with Microsoft Academic Graph citations:** Using data from the REF 2014 and citations from Microsoft Academic Graph, a study found relatively high correlations between departmental REF Grade Point Average output rankings and citation data in Chemistry (0.802) and Biological Sciences (0.797) (Pride & Knoth, 2018).

The above studies tended to emphasise the potential for bibliometrics to replace or supplement peer review in the REF or RAE, rather than the limitations, such as funding shifts between institutions if the scores (rather than rankings) change and the potential for perverse incentives when there is a financial incentive to achieve high bibliometric scores (e.g., moving away from less cited important research topics).

## 3.2. Peer review and bibliometrics in other countries

**Evidence from Australia:** Australia has used journal rankings decided by peer review to inform its national research evaluation, Excellence in Research for Australia (ERA). Although an early investigation found insufficient evidence of an association between citation-based journal metrics and the four tier ERA rankings of Australian social science journals (Haddow & Genoni, 2010), a medium degree of similarity was later found between three journal citation-based indicators and the expert-based ERA rankings. The Source-Normalised Impact per Paper (SNIP) had the highest Spearman correlation (0.54) with ERA rankings (n=11,137), followed by raw impact per paper (0.38) and the Journal Impact Factor (0.37) across 27 Scopus subjects, although there were some disciplinary differences. For instance, in Dentistry, journal-based citation metrics had the highest correlations with REA expert journal rankings (0.73, 0.78 and 0.72 respectively), followed by Chemical Engineering, and Veterinary Science, whereas very weak associations were found for Social Sciences (0.41, 0.24 and 0.26) (Haddawy et al., 2016).

**Evidence from Italy:** Italy uses an output-based periodic research assessment, known as the VTR and then the VQR. An investigation of institutional aggregate peer review ratings for academic publications submitted to the VTR and the journal impact factors of those publications found significant medium Spearman correlations for Biology (0.48), Chemistry (0.45) and Economics (0.44), suggesting that there is some degree of similarity between peer review outcomes and journal impact in some fields at the level of institutions (Reale et al., 2007). A large multidisciplinary study of over 12,000 research articles across ten subjects also found significant medium-high Spearman correlations between institutional aggregate peer ratings from Italian research assessment exercise and institutional aggregate article citations across most fields, such as Physics (rho=0.81), Earth Sciences (0.79), Biology (0.69), Chemistry (0.6) (Franceschet & Costantini, 2011). Another study found some agreement between citation indicators and VQR peer review ratings for 590 Italian articles in Economics, Management and Statistics (Bertocchi et al., 2015) and there has been an argument that that bibliometrics are preferable to peer-review due to cost savings from the time to perform peer review for Italian research assessment (see Abramo, D'Angelo, & Caprasecca, 2009; Abramo & D'Angelo, 2011). Nevertheless, recent evidence from the Italian research assessment exercise found that bibliometrics and peer review had weak associations in science, technology, engineering, and mathematics (Baccini, Barabesi, & De Nicolao, 2020).

**Evidence from the Netherlands:** The Netherlands does not have a periodic national REF-like procedure but has alternative methods of assessing research quality, sometimes using bibliometric indicators to inform expert judgement. An early investigation of 56 condensed matter physics programmes in the Netherlands found that in general there were positive relationships between a range of publication and impact indicators with peer judgements made by expert physics committees, although the strongest Spearman correlations were found between overall jury ratings and the average number of citations per publication (ranging from 0.51 to 0.68) and the field normalised citation averages (0.46 to 0.58) (Rinia et al., 1998). A later study of journal articles from 147 university chemistry research groups in the Netherlands (1991-2000) found that both the h-index and the 'crown indicator' (field normalised citation count) for research groups significantly and positively correlated with peer judgments of the research quality of published research (Van Raan, 2006).

**Evidence from the Norway:** A case study of 34 research groups from a Norwegian university found significant, albeit weak, correlations between expert panel ratings and various citation metrics, including relative subfield citedness (r=0.46), relative citation rate (0.24) and number of citations per person (0.31) (Aksnes & Taxt, 2004). There are also positive associations between different journal citation indicators (SNIP, Scimago Journal Rank and the raw impact per paper) and Norwegian expert-based assessments of journals and series (Ahlgren & Waltman, 2014).

# 4. Assessing the accuracy of score predictions for individual documents

The best way to assess the accuracy of AI predictions of quality scores for individual documents seems to be to compare them with expert human judgements, assuming these judgements to be correct. This section briefly reviews reasons why the judgements may be incorrect, before assessing the prediction accuracy of AI predictions achieved so far.

The quality of academic research is a subjective quantity that is often treated informally or, if defined, usually encompasses three dimensions: rigour, originality, and (scholarly and societal) significance (Aksnes, Langfeldt, & Wouters, 2019; Langfeldt et al., 2020). Each dimension is subjective and varies greatly between fields. For example, in the humanities, rigour applies primarily to argumentation and might entail a reasonably exhaustive consideration of evidence, possibilities and alternatives, together with convincing assessments of a variety of evidence sources. Qualitative methods rigour might focus instead on ethical dimensions of human subjects research, and the procedures used to tease themes out of data and understand the likely subjective influences of the author(s). From a technological perspective, construction engineering rigour might include the need for bricks to be baked in the appropriate type of oven. In many fields, rigor probably also involves using suitable statistical tests appropriately. Whilst mistakes are easy to identify in these contexts, it is more difficult to judge between levels of rigour for methods/approaches that are broadly appropriate. The originality dimension is clearly subjective. It depends on what the evaluator is already aware of and could be applied to different aspects of research (methods/approaches, research objects, objectives). Research significance in some specialties might be reasonably assessed with citation counts, but usually encompasses societal impact and evaluators are unlikely to have sufficient knowledge to reliably judge the extent of societal impact of a study, given the myriad potential impacts and the fact that non-academic pathways to impact are rarely documented.

A second issue is that quality can be judged from different perspectives, giving different outcomes (Langfeldt et al., 2020). In particular, work that is judged to be high quality within a field because it contributes to the internally agreed field goals may be less highly regarded in national research evaluations because the field goals are not known or are rejected, for example because they are judged to insufficiently consider societal perspectives by being too theoretical or methodologically problematic.

The above mainly generic problems assessing article quality are complicated by disciplinary differences in the extent to which the quality of an article can be reliably assigned, in the sense of different experts having a high probability of giving the same score. There are several reasons for this. First, there are differences in the extent to which fields are externally-focused, making research significance more difficult to assess. Second, there are differences between fields in the ease with which rigour can be assessed, due to standardisation of procedures or the lack of this (Barker & Pistrang, 2005). More generally, not all fields have a relatively uniform centralised agreement on what constitutes high quality research (Trowler, 2014). For example, whilst this might be expected from fields organised as conceptually integrated bureaucracies (Whitley, 2000) because of relatively centralised control of reputation allocation, it does not occur for fields with varied objects, objectives and/or methods (dis)organised as fragmented adhocracies (Whitley, 2000). In some senses in between these are polycentric oligarchies (Whitley, 2000), where quality is contested between warring paradigms, such as qual v. quant or empirical vs. theoretical. Other factors being equal, a much higher rate of agreement on quality scores would be expected from the first of the three organisational types.

Given the above factors affecting human judgements of article quality, imperfect human agreement can be expected for all academic fields and substantially different rates of human agreement between fields. These affect the maximum accuracy that it is achievable for AI systems: if the humans disagree on what constitutes quality, then it is more difficult for AI to learn from their decisions. In addition, if there are large disciplinary differences in the variety and standardisation of methods, objects and objectives within a field, then it is technically harder for AI systems to learn markers of quality because they are more diverse: the patterns to discover are fainter. For example, in health-related fields where randomised control trials are reasonably common and recognised as the most robust method, the AI can be expected to learn this. In contrast, most other fields probably do not have a single named high-quality method so it would be more difficult for the AI to distinguish a quality hierarchy of methods, if there is one. For all these reasons, little can be deduced by comparing AI system accuracies between fields. With this caution, accuracy statistics for AI (including statistical approaches with different training and test sets) in different fields is summarised below.

It seems that no previous published studies have used machine learning to predict the quality scores of individual articles, although it has been used to predict long term citation counts for individual articles and statistical methods have been used to predict quality profiles for sets of articles. The closest to a prediction of article-level quality scores was a set of threshold-based tables applied across all disciplines in a year of REF2014 data to predict whether an article had a 4* score or not (HEFCE, 2015). Although not the purpose of this test, the data can be converted into accuracy statistics and compared to a baseline strategy of predicting that no articles are 4*. From this comparison, it is not surprising that all strategies had negative accuracy compared to the baseline, although raw citation count strategy was closest to achieving a positive result (Table 3). Because of the disciplinary differences mentioned above, this simple strategy could have achieved a positive accuracy above the baseline for some UoAs.

Table 3. Accuracy statistics for article-level predictions of whether a REF2014 journal article from 2008 had a 4* score or not across all 36 UoAs (calculated from the two-way summary tables in: HEFCE, 2015). The baseline is predicting that no article is 4*.

| Indicator | Accuracy | Baseline | Accuracy above baseline | Articles |
|---|---|---|---|---|
| Scopus citation counts | 76.4% | 76.6% | -0.2% | 21060 |
| Google Scholar citations | 76.2% | 76.4% | -0.2% | 21055 |
| FWCI (field normalised citations) | 75.4% | 76.1% | -0.7% | 19580 |
| Highly cited percentiles | 79.3% | 91.1% | -11.8% | 19675 |
| SNIP (a field normalised JIF variant) | 74.6% | 76.2% | -1.6% | 19130 |
| SCImago journal rank | 74.7% | 76.1% | -1.4% | 19245 |
| WIPO patent citations | 76.1% | 96.9% | -20.8% | 21060 |
| Mendeley readers | 74.9% | 86.4% | -11.5% | 21050 |
| ScienceDirect downloads | 67.2% | 76.0% | -8.8% | 6990 |
| Scopus full text requests | 68.3% | 76.2% | -7.9% | 21060 |
| Tweets | 74.7% | 94.2% | -19.5% | 21055 |

# 5. Availability of relevant public datasets

REF automation may be supported by public datasets or sources of bibliometric information, such as field normalisation scores, and large altmetrics databases such as Dimensions or that provided by Altmetric.com to researchers (see also section 9 for open review databases). This section will survey these, as well as relevant public APIs for data harvesting, such as COCI (Crossref's OpenCitations Index).

## 5.1. Sources of open citations

The coverage of conventional citations indexes may not be sufficient for the wider impact assessment of research, especially in the arts and humanities or social sciences (Moed, 2005, see also below). Moreover, traditional citation indexes like the Web of Science and Scopus may not fully reflect the citation impact of recently published or in press articles (Kousha, Thelwall, & Abdoli, 2018) and hence other open citation platforms and academic search engines could be helpful for timely research evaluation to identify research that quickly attract many citations. Although traditional citation indexes seem to index citations faster now, such as through in press citation indexing, they are less comprehensive than other sources.

## 5.2. Google Scholar citations

Google Scholar is a free search engine of online scholarly publications such as articles, theses, books, and conference papers. It indexes publications from many sources, such as academic publishers, preprint or postprint repositories, grey literature and other websites (e.g., web CVs, university archives).

**Google Scholar has a wider coverage of scholarly-related publications than traditional citation indexes:** Many early small-scale investigations have compared the coverage and citation statistics of Google Scholar against Web of Science or Scopus, finding that Google Scholar had wider coverage of academic publications and found more citations (e.g., Meho & Yang, 2007; Kousha & Thelwall, 2007; Bar-Ilan, 2008; Kulkarni et al., 2009; Mingers & Lipitakis, 2010; De Groote & Raszewski, 2012; de Winter et al., 2014). Very high correlations have found between Google Scholar citation counts and Web of Science or Scopus citation counts across many subject areas (for a review see Appendix A in Thelwall & Kousha, 2015a). An early study estimated that Google Scholar had indexed 87% (100 million of 114 million) of all English-language scholarly documents on the web (Khabsa & Giles, 2014; broadly agreeing with: Aguillo, 2012). Another study estimated that in May 2014 Google Scholar had three times more scholarly records than the Web of Science (171 million compared to 57 million) (Orduña-Malea et al., 2015). More recently Google Scholar was estimated to index 389 million scholarly related records, substantially more than Scopus (72.2 million) and the Web of Science (67.7 million) (Gusenbauer, 2019). A large-scale comparison of Google Scholar, Web of Science, and Scopus across many subject areas found that there were very high Spearman correlations (mostly close to 1.0) between Google Scholar citations with either Web of Science or Scopus citation across 252 specific subject categories, except Literature (0.78). Google Scholar found the largest percentage of all citations found across all subject areas (ranging from 93% to 96% depending on the area) compared with Scopus (35%–77%) and WoS (27%–73%). In most social sciences and art and humanities subjects, Google Scholar found citations outside both Web of Science and Scopus, mostly from non-journal materials (e.g., dissertations, books or book chapters, conference proceedings or preprints) (Martín-Martín et al., 2021).

**Potential application of Google Scholar citations in the UK Research Excellence Framework:** Google Scholar could be a source of citation counts when citations from a broad range of international publications (especially non-English) or recently published research is required, especially in the arts and humanities with relatively low coverage in traditional citation indexes (see above studies). In the Sub-panel 11 (Computer Science and Informatics) of REF 2014, Google Scholar was recognised as helpful additional citation source, "where outputs have been cited extensively outside the body of publications indexed in Scopus[2]". However, because Google Scholar does not support large-scale automatic searches and manipulation of citation counts is easy (Beel & Gipp, 2010; López-Cózar, Robinson-García, & Torres-Salinas, 2014), it is problematic to use for AI-assisted research assessment exercises.

## 5.3. Google Books Citations

Google Books is not a citation index but citations from its digitised books and monographs could be another potential source for citation impact assessment of book-based fields, which otherwise lack good sources of impact data. A study of 3,573 journal articles in ten science, social science and humanities subject areas found that Google Books citations were 31%-212% as numerous as Web of Science citations in the social sciences and humanities, but were relatively rare in the sciences (only 3%-5%). The study also found quite high correlations between Google Books and Web of Science citation counts in all subjects, although this association was higher in computer science (rho=0.709, perhaps to conference proceedings), philosophy (.654) and linguistics (.612) than in chemistry (.345) and physics (.152) (Kousha & Thelwall, 2009).

**Potential application of Google Books Citations in the UK Research Excellence Framework:** A study of 1,000 books submitted to the RAE 2008 in seven book-based subjects (archaeology, law, politics and international studies, philosophy, sociology, history, and communication, cultural and media studies) found that Google Books citations to books were 1.4 times more frequent than Scopus citations. In history, for instance, the median number of Google Books citations (11.5) was higher than for both Google Scholar (7) and Scopus (4) citations (Kousha, Thelwall, & Rezaie, 2011). This suggests that Google Books could be useful consulting source for scholarly impact assessment of book-based fields, where the coverage of traditional indexes is not sufficient and many articles may be cited in books rather than or in addition to articles. Although it is possible to automatically search for citations to most articles using the Google Books API with relatively high accuracy and coverage, the method may return false matches for articles with very general or short titles (Kousha & Thelwall, 2015a). Hence, the automatic Google Books citation extraction method for AI-assisted research assessment exercises could be problematic to be used for all publications in some cases may need extra manual checks. Another issue is that it is not known whether book-to-book citation counts tend to reflect the quality of books in any arts and humanities subject areas.

## 5.4. Dimensions

The scholarly search engine Dimensions (dimensions.ai) is similar to Google Scholar, but has more sophisticated search functions and is has the document categorisation capability that it useful for effective citation impact assessment. Its core features are free but it incorporates some paid services and is owned by Digital Science. It integrates with other databases, such as for funding and patents. Dimensions claims to include "more than 126 million publications from 93,000 journals, 64 preprint servers and over 1 million books" in addition to links to other records such as millions of patents, clinical trials, datasets, policy documents and supporting grants[3].

[2] https://www.ref.ac.uk/2014/media/ref/content/pub/panelcriteriaandworkingmethods/01_12_2B.doc

**Dimensions has wider coverage of scholarly documents than Scopus and WoS:** An initial study found that the coverage of Dimensions was comparable to that of Scopus (97%) and there was a high correlation between citation counts from Scopus and Dimensions (rho=0.96) (Thelwall, 2018b). A later paper about library and information science (journal, document and author levels) also found a very strong association between Scopus and Dimensions citation counts (rho=0.96) and that Dimensions coverage of the recent literature is similar or slightly better than Scopus but less than Google Scholar (Orduña-Malea & López-Cózar, 2018). A report about the six top journals in Business & Economics also found that that Dimensions is a more comprehensive source than Scopus and the Web of Science for locating relevant research and citation analysis and has similar coverage to Crossref but not as complete as Google Scholar and Microsoft Academic (Harzing, 2019). Another large investigation compared Web of Science master journal lists (13,610 journals) against the journal coverage of Scopus and Dimensions, finding 99.1% and 96.6% overlap with them, respectively. Scopus indexed journals also had a 96.4% overlap with Dimensions. It also compared the publication records from 20 countries 2010-2018 across three databases, finding that Dimensions had about 82% and 48% more indexed journals than Web of Science and Scopus respectively (Singh et al., 2021). A very large-scale comparison of five citation sources (Web of Science, Scopus, Dimensions, Crossref, and Microsoft Academic) found that Dimensions and Crossref had similar coverage of scientific documents from 2008–2017 (36 and 35 million records, respectively) and higher than both Scopus (27 million) and Web of Science (23 million). Microsoft Academic had the largest database, with 73 million documents (Visser, van Eck, & Waltman, 2021), although it is not accessible anymore[4].

**Dimensions' (early) field classification scheme may not be very accurate:** Dimensions uses an AI/machine learning based approach to automatically categorise publications. A study of 262 publications by an individual researcher active in scientometrics, informetrics, bibliometrics, and altmetrics found that most articles were misclassified such as ''Applied Economics'' and ''Public Health & Health Services'' (Bornmann, 2018). It is an evolving system that appears to have greatly improved since its early versions, although there do not seem to be empirical studies to verify this.

**Potential application of Dimensions in the UK Research Excellence Framework:** Dimensions has several API services to perform searches and analysis, hence could be a potential source for large-scale research evaluation exercises (Herzog, Hook, & Konkiel, 2020). However, like Google Scholar, its coverage from indexed peer reviewed sources is not fully clear and may change substantially over time and seems to be mostly dependent on data from Crossref (see Visser, van Eck, & Waltman, 2021). There is evidence that about half of Dimensions indexed records do not have affiliation countries (Guerrero-Bote et al., 2021) and field classification of publications has been imperfect (Bornmann, 2018).

[3] https://www.dimensions.ai/

[4] https://www.microsoft.com/en-us/research/project/academic/articles/microsoft-academic-to-expand-horizons-with-community-driven-approach/

## 5.5. OpenCitations

The COCI dataset, (the OpenCitations Index of Crossref open DOI-to-DOI citations) is the first fully open scholarly bibliographic and citation dataset. It is contributed to by publishers as a common resource. By April 2022, COCI contained 1.3 billion citations and 72 million bibliographic resources[5]. Because of low quality metadata in other large open citation platforms, such as Google Scholar and the retirement of Microsoft Academic in December 2021, the OpenCitations project could be a significant platform to access open bibliographic and citation data which is downloadable in web standard Resource Description Format (RDF) (see also Heibi, Peroni, & Shotton, 2019; Peroni & Shotton, 2020). OpenCitations could theoretically be a potential source for evaluators, funders or national research assessment exercises to assess the wider citation impacts of research in a timely manner. However, a large-scale study of over 3 million citations to 2,515 English-language highly-cited publications in 2006 retrieved by six citation databases found that OpenCitations' COCI was the smallest, with 28% of all citations compared with Google Scholar (88%), Scopus (57%), Dimensions (54%) and Web of Science (52%), suggesting that public citation data was not large compared with other sources at the time of study (Martín-Martín et al., 2021). With the recent addition of extra citation data to COCI (1,294,283,603 citations, see http://opencitations.net/index/coci) this difference in citation coverage might be smaller.

## 5.6. Sources of alternative metrics

### 5.6.1. Social media indicators

The altmetric data providers Altmetric.com, PlumX and others capture mentions of scholarly publications in social media platforms (e.g., Facebook, Twitter, Reddit, or Blogs) or scholarly related sources (e.g., Mendeley, Wikipedia, Faculty Opinions, or Patents). Crossref Event Data and Mendeley.com also provide API services to capture mentions of publications in their online platforms. These have been proposed as sources of impact evidence to reflect societal or other specific impacts, often to complement citation counts.

There are differences between altmetric platforms in terms of their coverage of metrics and publications (Ortega, 2020; Karmakar et al., 2021). Many early studies showed that alternative metrics from social media sites significantly correlated with bibliometric indicators (e.g., Priem, Piwowar, & Hemminger, 2012; Thelwall, Haustein, Larivière, & Sugimoto, 2013; Costas, Zahedi, & Wouters, 2014; for reviews see: Thelwall & Kousha, 2015b; Sugimoto et al., 2017). There is evidence that among all altmetric sources, Mendeley reader counts have the strongest correlations with citation counts across many fields (Thelwall, 2017b) and can be helpful indictors to predict the future citation impact of research (Thelwall, 2018c; Thelwall & Nevill, 2018). They might theoretically play this role in future AI-assisted research assessment exercises if they were not easily gamed. However, an independent review of the role of metrics in research assessment and management in the UK stated that "although alternative metrics do seem to give indications of where research is having wider social impact, they do not yet seem to be robust enough to be routinely used for evaluations in which it is in the interest of stakeholders to manipulate the results" (Wilsdon et al., 2015b, p. 49). This statement is equally true in 2022. The problem is that almost all altmetric indicators can be easily manipulated (Rasmussen & Andersen, 2013) if they are used in research assessment. Although it seems that manipulation of

[5] http://opencitations.net/index/coci

Mendeley readers is more difficult than other altmetric indicators, it is possible for authors to ask other Mendeley users (e.g., students or colleagues) to register their articles or use other methods to increase their reader counts. Nevertheless, altmetric indicators could be useful to identify non-academic benefits of research for impact case studies within the REF, where societal impact claims might be evidenced mainly through non-scientific sources, such as news or social media posts. For instance, there is evidence that publications cited in the REF 2014 impact case studies tended attract more web mentions (Twitter, Wikipedia, Facebook, blogs, news, and policy-related documents) than submitted REF research outputs (Bornmann, Haunschild, & Adams, 2019) and there is an association between altmetric scores and expert peer review ratings of publications referenced in REF 2014 impact case studies (n=1,469) submitted under main panel B (Wooldridge & King, 2019).

**Mendeley reader counts moderately correlate with REF quality scores in Clinical Medicine and Biological Sciences:** There seem to be only two studies about associations between altmetrics and peer-review outcomes. A large-scale study of REF 2014 outputs found overall significant but very low correlations between REF 2014 peer review scores and Mendeley reader counts at the article level (rho=0.19). This association was higher in Clinical Medicine (0.441) and Biological Sciences (0.363) than in other subjects (HEFCE, 2015, Table A39). Using a regression analysis, another study found that among several studied factors only citation counts and Mendeley readers significantly correlated with quality scores on the F1000 platform (Bornmann & Haunschild, 2018).

### 5.6.2. Other Sources of Online Impact

A range of specialist online sources could be useful for the wider impact assessment of research in particular cases. These include citations in clinical trials or clinical guidelines (Thelwall & Kousha, 2016; Thelwall & Maflahi, 2016), digitised patents (Kousha & Thelwall, 2017), grey literature publications (Bickley, Kousha & Thelwall, 2021) and books and non-standard outputs (Kousha & Thelwall, 2015b). These sources seem to be more useful for registering non-academic benefits of research, such as for REF impact case studies (Kousha, Thelwall, & Abdoli, 2021), rather than for the individual impact assessment of articles because they are rare. In addition, many types of non-academic impacts cannot be easily captured through current databases and may need extensive web citation searches to identify, which might not be feasible for large-scale research evaluation exercises (for a review see: Kousha, 2019).

# 6. Transparency in technology assisted assessment

Transparency in technology assisted assessment has been argued to be important to allow those assessed to check the results and suggest corrections for mistakes, if necessary. One of the ten principles of the Leiden Manifesto for research evaluation is, "Keep data collection and analytical processes open, transparent and simple" (Hicks et al., 2015). The simplicity aspect runs against AI processes, which are usually complex, and its objective is to support transparency by allowing those evaluated to understand the processes involved enough to check then. This also aligns with the open science agenda to make all aspects of science available for inspection (Bornmann et al., 2021). Different relevant aspects of transparency are discussed here as well as relevant AI considerations.

## 6.1. Transparent data sources and processing

Bibliometric data sources are mostly controlled by commercial organisations such as Dimensions.ai (Digital Science), Scopus (Elsevier) and the Web of Science (Clarivate). These organisations broadly publish their methodologies for finding and including journal articles. The main sources are manually curated lists of academic journals for Scopus[6] and the Web of Science[7], which are published and public. The process of choosing these journals is human-based and private, although the outcome is public. The procedure used to classify journals into field-based categories (Scopus, Web of Science) also seems to be manual but probably helped by automated analyses of the references and citations of each journal. Journal classification is an important aspect of non-transparency because a journal's categories can have a substantial influence on whether its articles tend to be cited above or below the world average for its categories.

Elsevier and Clarivate presumably have agreements with the publishers to harvest relevant information about the journals from the publishers' websites and then use their own private algorithms to transform the raw data into bibliometric information. These algorithms would include those that use simple heuristic rules or a form of AI for non-trivial tasks, such as the following:

- Matching reference lists to cited documents in the absence of DOIs.
- Disambiguating author names for search functions that identify an individual researcher's works.
- Matching affiliations to author names.
- Classifying articles by field for field normalisation purposes (Clarivate only).
- Identifying multiple copies of the same publication.

The first of these is the most important for research evaluation since errors in reference matching can reduce citation counts (e.g., Harzing, 2017; van Eck & Waltman, 2019), and duplicate publications can cause the same problem by sharing citations (van Eck & Waltman, 2019). Errors can originate from many minor sources, and the varied algorithms used means that it is not possible to publish transparent versions that can be checked. Nevertheless, there does not seem to be a simple way to request correction of referencing errors in these databases.

Dimensions.ai uses public Crossref data provided freely by publishers as well as arrangements with other publishers to directly harvest their bibliometric metadata, and crawlers to harvest various repositories, such as PubMed and arXiv[8]. It does not publish a list of journals indexed but explains how to check if a journal is indexed[9]. Its processing transparency issues are similar to those of Elsevier and Clarivate but has an additional source of algorithmic opaqueness: the AI algorithm used to classify articles into fields. This is only a minor issue since the human decision making of Clarivate and Elsevier for journal classification is similarly opaque.

[6] https://www.elsevier.com/solutions/scopus/how-scopus-works/content
[7] https://mjl.clarivate.com/search-results
[8] https://plus.dimensions.ai/support/solutions/articles/23000018860-how-is-the-publications-data-harvested-

## 6.2. Transparent research indicator calculations

The indicators reported by bibliometric databases seem to be relatively transparent in the sense that the formulae are published and tend to be simple and checkable. This strategy presumably helps scientist to understand and adopt them. Clarivate[10], Elsevier[11] and Dimensions[12] publish and explain the indicators used. A few of the metrics, such as SCImago Journal Rank, are not transparent because they rely on large matrix factorisations that integrate the entire bibliometric database in one high-dimensional matrix calculation.

## 6.3. Transparent AI

Away from the field of research evaluation, AI researchers have addressed the problem of most machine learning algorithms being opaque in the sense of being too complex for an intuitive understanding of how they work in a particular case. For example, a deep learning model may be a neural network with thousands of interconnected nodes, with each connection having its own weights. Whilst the input and output layers may be interpretable, the intermediate layers may not have an intuitive understanding even if there were not too many nodes to follow anyway. Similarly, Support Vector Machines operate in high dimensional spaces that are beyond human understanding. In contrast, the decision tree is a simple algorithm that is easy to understand because it requires checking multiple transparent decisions. Three current state-of-the-art machine learning algorithms, random forest, gradient based classifier, and extreme gradient boost, all use hundreds of simultaneous decision trees, combining them using mathematical formulae for the output (Chen et al., 2015). Thus, although their building blocks are transparent, the algorithms overall are not.

Algorithmic opaqueness makes it more difficult to check that an algorithm has not introduced biases and makes it more difficult for the algorithm owner from being accountable for decisions (e.g., Diakopoulos & Koliska, 2017). This has led to the field of eXplainable AI (XAI) or “white box” AI (Vilone & Longo, 2020; Xu et al., 2019), which focuses on algorithms with decision making process that a human expert could understand, such as linear regression, a finite set of rules, or a decision tree. This might also allow a specialist to adjust part of the AI based on their knowledge that it was incorrect even though it was consistent with the dataset that the AI had been trained on (Gunning et al., 2019). There are different grades of transparency in XAI, with the most transparent being explainable to end users rather than AI experts.

## 6.4. Transparent AI and the REF

End user understanding may be important to give confidence in a system, such as in the context of REF-related AI. Ideally, any REF-related AI would be fully transparent so that researchers could verify all aspects of the input and understand all the steps that the algorithm used to get the answer (e.g., a score or recommended assessor for an output). This would typically sacrifice accuracy, however, since state of the art algorithms are not transparent for most machine learning tasks.

[9] https://www.dimensions.ai/submit-journal-and-book-titles/
[10] https://clarivate.com/webofsciencegroup/wp-content/uploads/sites/2/2021/06/JCR_2021_Reference_Guide.pdf
[11] https://www.elsevier.com/solutions/scopus/how-scopus-works/metrics/citescore
[12] https://plus.dimensions.ai/support/solutions/folders/23000031268

To set the above discussion in context, the human experts in the REF that make the key decisions, such as selecting panel members to assess outputs and assigning a score to outputs, are also not transparent. In particular, panel members use their subject expertise, knowledge of the REF rules and discussions with other panel members to reach their decision. It seems likely that many of the decisions about scores are reached based on intuition with a component of emotional reaction, "is this research exciting", rather than through simple explainable processes, especially in higher numbered UoAs. In any case, the decision-making process is not communicated to the output authors and so is 100% opaque. Instead, authors are not told who evaluated their outputs and are given vague feedback about large sets of outputs, such as "[within the set of 100 outputs submitted] those on the topic of [x] were considered particularly strong". The decision to hide individual output scores and give no feedback about them in the current REF makes it unlikely output authors would be communicated their results with transparent AI, but transparency would be an advantage for panel members seeking confidence in AI system outputs.

# 7. Sources of bias in technology assisted assessment

Bias is an "inclination or prejudice for or against one person or group, especially in a way considered to be unfair"[13]. In the context of the REF, biases might be against individual people, institutions, research methods, genders, career stages, output types, or negative findings, for example.

## 7.1. Algorithmic bias

Algorithms can show bias and make biased decisions (Kordzadeh & Ghasemaghaei, 2021; Mehrabi al., 2021; Navarro et al., 2021), as illustrated by some high-profile cases. For example, a recruiting tool from Amazon was shelved after it was shown to be biased against women[14]. AI systems can be biased because they are fed biased rules, learn from biased data, or accidentally introduce bias as a side-effect of something else. There are different types of algorithmic bias.

**Design bias**: This can occur if a system is poorly designed. For example, a facial recognition system that is only trained on white faces because of the prejudice or thoughtlessness of its creators would be biased (Furl et al., 2002; Lee, 2018) and this could have unpleasant effects when it is used in practice. Alternatively, an inappropriate set of inputs to a system might be selected so that it is not shown important information because the designers did not realise its value. For example, an AI system to estimate the quality of candidates based on their career achievements would be biased against women if it was not fed career gap information.

**Existing bias**: The system learns existing prejudices in society from its input data and conforms to them. For example, since some job categories are heavily gendered (e.g., nurse, carpenter), a machine learning system designed to recommend jobs to candidates based on their CVs could easily learn and then exacerbate existing gender divisions by only recommending carpentry to men and nursing to women. Such an algorithm might also primarily recommend senior jobs to men, or lower paid jobs to ethnic minority candidates. Here the system notices a pattern (e.g., most previously interviewed candidates for top jobs have been male) and then uses the gender on a CV, together with other information, to help predict whether the person should apply for a senior role. Whilst women and nonbinary people might still be recommended to apply, on average, their CVs would have to be better to trigger this recommendation.

**Indirect bias**: An AI system makes biased decisions because of factors unrelated to its primary design goals. For example, a system paying to show adverts to users of an electronic system might primarily target the cheapest demographic to reach the largest audience. This might lead to career adverts disproportionately targeting the cheapest gender (Lambrecht & Tucker, 2019) or age group unless the system is configured specifically for demographic equality. Similarly, sentiment analysis systems have been shown to disproportionately reflect the opinions of demographics that express sentiment most clearly, such as women compared to men (Thelwall, 2018a).

## 7.2. Bibliometric bias

When bibliometric data is used to support assessment then there is the potential to introduce many types of bias. The main ones are summarised here. Although gender bias is widely believed to occur (e.g., Rowson et al., 2021), this is not an issue from the REF perspective because female first-authored articles tend to be slightly more cited than male first-authored articles in the UK (Thelwall, 2020). This is counter-intuitive because men typically dominate citation-based lists based on career citations or the h-index. This domination tends to happen because men tend to have fewer career gaps, are less likely to leave academia, and retire later. Because of these factors, they tend to accrue more career citations. In addition, today's older academics started when there were larger obstacles to women entering academia than there are today.

---

[13] https://www.lexico.com/definition/bias

[14] https://www.reuters.com/article/us-amazon-com-jobs-automation-insight/amazon-scraps-secret-ai-recruiting-tool-that-showed-bias-against-women-idUSKCN1MK08G

**Field biases**: Academic fields cite at different rates, with different length reference lists, citing different balances of journal articles, books and other outputs, and citing different age outputs. Because of this, average citation counts differ substantially between fields. Citation counts should therefore only be compared between articles from the same field, unless field normalised or percentile indicators are used instead (Thelwall, 2017a). This also applies to Journal Impact Factors, which should not be compared between fields. Of course, citation counts should also not be compared between articles of different ages, unless with field normalised or percentile scores.

**Research type biases**: Some types of research are naturally more cited than others, which introduces another citation bias. Review papers are the clearest example of a type that is usually more cited (Aksnes, 2003). Articles using particular methods can also tend to be more highly cited (Antonakis et al., 2014; Fairclough & Thelwall, 2022; Thelwall & Nevill, 2021). In particular, it seems likely that, within mixed methods fields, papers making more hierarchical contributions (e.g., incremental method improvements) or contributing to faster publishing specialisms (e.g., simulation modelling rather than interview-based studies) will tend to be more cited, or at least cited more quickly. Papers in an expanding research area are also likely to be more cited because there are relatively many citing papers compared to the number of potentially cited papers. Positive results are also more likely to be cited (e.g., Jannot et al., 2013; Tincani & Travers, 2019; Urlings et al., 2021), although in REF terms these might also be judged to be more significant.

**Country biases**: Citation bias is likely against research from, about, or in the languages of, countries that are not well indexed in the bibliometric database used for a citation analysis. This is particularly likely for domestic issues. All major citation databases make decisions about which journals to cover and they seem to primarily cater for English-language searches so this leads to a bias against research that is from countries where research is often not written in English (Mongeon & Paul-Hus, 2016; van Leeuwen et al., 2001). Since researchers are disproportionately cited from their own country (Lancho Barrantes, et al., 2012; Thelwall & Maflahi, 2015), under-indexing the work of a country creates a citation bias against the articles that are indexed. This is exacerbated for nationally-focused research that would expect to rarely be cited from other countries, perhaps including studies on indigenous plants and animals. From the REF perspective, the UK is well indexed by all major databases but academics with interests that focus on less well-indexed countries (e.g., some Area Studies) may be disadvantaged.

**Research volume bias**: Related to country biases, an article on a topic that few researchers are publishing about will tend to be less cited than articles on popular topics. This may be legitimate if the more researched topic is more important but not legitimate if the topics are equally important but there is more activity about one topic for economic reasons. For example, an ecological researcher in a region with a relatively unusual characteristics and few researchers may be rarely cited for this reason (Culumber et al., 2019).

**Recognition/prestige bias**: Researchers may prefer to cite work from well-known people (the "Matthew effect", Merton, 1968), or prestigious sources (journals, institutions) because they are biased in its favour or consider it to be a safe option. Well known works can also be cited as concept markers for a topic rather than for their contents (Case & Higgins, 2000).

## 7.3. Peer review bias

Several factors are known to influence peer review decisions, as summarised relatively recently (Lee et al., 2013). These biases are also likely to translate into citation biases when academics review articles when deciding whether to cite them. It is known that even the most expert academic peer reviewers sometimes make poor decisions, such as editorial rejection of important articles (Siler et al., 2015), but this section focuses on systematically sub-optimal decisions with an identifiable cause.

**Prestige bias**: Reviewers may form more favourable judgements for outputs from successful researchers (Merton, 1968; Tran et al., 2020), from more prestigious institutions, or for articles that they believe are standard to cite in the field (Brooks, 1986).

**Nepotism**: Academic reviewers may form more favourable judgments of the work of people that they know (Sandström & Hällsten, 2008).

**Gender bias**: Although universities have historically been extremely sexist institutions, there is not a consensus about whether gender bias in academic evaluations remains a problem. There is not strong empirical evidence of overall gender bias in judgements (Ceci et al., 2011) despite persistent problems with the underrepresentation of women in senior positions. Nevertheless, there are areas or aspects of science that are chilly climates for female researchers (Biggs et al., 2018).

**Nationality/ethnicity**: Reviewers may be prejudiced against the work of academics from particular countries or ethnicities (Hojat et al., 2003).

**Cognitive bias and distance**: This occurs when judgments are influenced by the reviewers' beliefs about the subject matter without considering whether their beliefs are universal (e.g., Bader et al., 2021). This can occur in two ways: a researcher from a distant field may undervalue a study through a lack of understanding of its importance, or a researcher from a competitive paradigm may not value a study at all. Although empirical evidence in limited contexts shows the opposite of what might be predicted that reviewers are stricter on topics closer to their own area (Boudreau et al., 2016; Wang & Sandström, 2015), variations of cognitive bias seem likely to be widespread or universal and unavoidable, at least in the first form. Cognitive distance presumably applies to all interdisciplinary research to some extent, since reviewers may be unfamiliar with some of the component disciplines (Rinia et al., 2001).

**Confirmation bias**: Closely related to the above, a reviewer may be more critical of work that challenges their beliefs (Mahoney, 1977).

**Novelty bias**: The most novel research can sometimes have difficulty in passing peer review and eventually be published in less prestigious journals than the subject merits (Campanario, 2009; Gans & Shepherd, 1994; Wang, Veugelers, & Stephan, 2017). **Layout bias**: Reviewers may be influenced by first impressions based on article layout (e.g., Moys, 2014). For example, if they review a preprint in an awkward format (e.g., double spaced, with figures and tables at the end) they may be more likely to give a negative evaluation than if they had read the journal printed version.

## 7.4. Bias in technology assisted assessment

A technology assisted assessment system that seeks to predict peer review scores by learning patterns associated with research quality from bibliometric and other data is likely to inherit some but not all of the biases of bibliometrics and peer review.

In terms of bibliometric inputs, higher citation counts associate with higher quality research to varying extents in most fields, so an AI system is likely to leverage citation counts. If it is fed field normalised citation counts rather than raw citation counts then this will avoid substantial biases against low citation fields. Even with field normalised specialisms, AI systems will still inherit biases against low citation types of research, as well as the country, prestige and research volume biases discussed above. Since the AI system will learn from peer review scores and assuming that the peer review scores did not reflect the same biases as the citations, then the AI system would, in theory, be able to learn to correct the AI bias with the human scores. In practice, this is unlikely to work perfectly because an AI system is unlikely to be fed with enough training data to learn any patterns reflected in a small minority of the article scores. Thus, depending on the volume of training data and the number of articles in the set that the bias is against, the bibliometric bias may be largely replicated by the AI or partially bypassed.

Some but not all peer review biases are also likely to be learned by an AI system that predicts quality scores and is trained on a set of journal articles with bibliometric information and peer review scores. This essentially depends on whether the relevant biasing information is fed into the AI system in the learning phase and the variety of the reviewer judgements. In the case of prestige information, if the AI system is not fed author career information, then it could not directly learn a prestige bias from the human reviewer scores and if it is not fed the gender and nationality of the authors then it cannot learn gender and nationality bias directly, even if it is present in the peer review scores. AI is also likely to ignore layout bias, as it would presumably not be fed with layout information.

A system could learn cognitive distance bias and confirmation bias if it dominated the peer review scores of relevant articles. For example, if all education reviewers gave low scores to qualitative research because they thought that quantitative research was inherently superior, then the AI system would probably learn to be biased against qualitative research. On the other hand, if the reviewers were evenly split between those that favoured quantitative and those that favoured qualitative research, then the AI system may well not learn a qual/quant bias and be less biased than individual reviewers in this regard. Similarly, if one topic was cognitively distant from all reviewers then the AI might learn to allocate lower scores to that topic.

Automatic translation systems can introduce gender biases (Prates et al., 2020) and so AI systems relying on translation (e.g., for articles not written in English and without an English translation) may introduce gender biases.

AI systems processing textual input as part of quality score prediction may generate biases against minority groups through language expression (Cheuk, 2021). For example, one empirical study has developed AI systems to predict conference review accept/reject decisions from word frequency text analysis of the submitted papers. The factors found most useful by the system were all superficial and indirectly associated with higher quality rather than measures of it: avoiding “quadratic”, few sentences, many difficult words, many pages, and many syllables per word (Checco et al., 2021). This approach seems likely to generate a bias against non-native English speakers who may prefer to use more straightforward language.

# 8. Field categorisations for journal articles

The method to automatically categorise journal articles into fields is an important aspect of field normalisation for AI systems predicting peer review. It is also relevant as a potential aid to sub-panel chairs assigning outputs to reviewers. The latter is a substantial time-consuming task that delays the start of the reviewing process.

Whilst the most well-known field classifications, from the Web of Science (WoS) and Scopus, are based on human classifications of journals, automatic methods are available to classify at the article level, as implemented by Dimensions.ai. WoS also has an automatic method to reclassify articles in multidisciplinary journals for some of its indicators (Clarivate, 2022). These methods typically exploit the references of articles, the citations to them and/or words or phrases in the full text or title/abstract/keywords. They can generate more accurate field classifications by clustering articles into thematically related sets or classifying them into pre-defined classes (e.g., Dimensions). These classifications can help with identifying reviewers for articles as well as making field normalisation calculations for citations or other bibliometric indictors (e.g., research collaboration) more accurate. This is one of the principles in The Leiden Manifesto for Research Metrics (Hicks et al., 2015). Different Unit of Assessment groupings might also be suggested, in theory. The subject classification of sciences used in citation databases (e.g., WoS, Scopus, Dimensions) can have a substantial impact on the field normalisation of scientometric indicators for research assessment exercises (Glänzel et al., 2009). Hence, there have been many studies about alternative classification systems for field normalised impact indicators (for reviews see Waltman & van Eck, 2019; Gläser, Glänzel, & Scharnhorst, 2017).

## 8.1. Journal-based field categorisations

Both Scopus and the Web of Science use field classification based on assigning academic journals to one or more subject categories. Hence, articles are categorised based on their journals' subject(s) rather than article topics. Web of Science uses 254 specific "Subject Categories"[15] in addition to 153 broader "Research Areas"[16] for journal classification and users can search a record by using advanced search commands "WC=" and "SU=" in these fields respectively. For instance, Web of Science has 7 subjects for chemistry journals: "Chemistry, Analytical", "Chemistry, Applied", "Chemistry, Inorganic & Nuclear", "Chemistry, Medicinal", "Chemistry, Multidisciplinary", "Chemistry, Organic", and "Chemistry, Physical". These subjects combine to form the "Chemistry" Research Area. Scopus uses 27 broad subjects in its main database and 334 narrow subject codes[17] which can be searched with the SUBJTERMS advanced search command (e.g., SUBJTERMS(2310) for articles assigned to Pollution). There are other journal-based schemes such as the Science-Metrix non-overlapping classification with six domains, 22 fields and 176 subfields[18] (see also Archambault, Beauchesne, & Caruso, 2011) or the DOAJ: Directory of Open Access Journals (https://doaj.org/) with 518 hierarchical subjects (e.g., Fine Arts/Architecture/Architectural drawing and design). Some researchers have also produced new classifications of science fields. For instance, a two-level classification scheme of science fields and subfields was proposed for research evaluation reasons, comprising 12 broad categories and 60 subfields for the sciences and 3 major fields and 7 narrow subjects of for both the social sciences and the humanities (Glänzel & Schubert, 2003). The UCSD (University of California, San Diego) Map of

---

[15] https://images.webofknowledge.com/images/help/WOS/hp_subject_category_terms_tasca.html
[16] https://images.webofknowledge.com/images/help/WOS/hp_research_areas_easca.html
[17] https://service.elsevier.com/app/answers/detail/a_id/14882/supporthub/scopus/~/what-are-the-most-frequent-subject-area-categories-and-classifications-used-in/
[18] https://science-metrix.com/classification/

Science classification system categorised Web of Science and Scopus journals into 13 broad fields and 554 subdisciplines (Börner et al., 2012).

There are conflicting subject classifications for some scientific journals. For instance, the *Journal of the American Medical Informatics Association* has been classified under the broad and narrow subjects "Medicine" and "Health Informatics" in Scopus, respectively, whereas Web of Science classified it under multiple Research Areas: "Computer Science", "Health Care Sciences Services", "Information Science Library Science", and "Medical Informatics". Science-Metrix classified this journal in the broad field "Information & Communication Technologies" with sub-field "Medical Informatics". Although the subject categories used in the above databases were mainly developed for information retrieval, they can influence bibliometric results. For instance, there is evidence that the set of journals in the Web of Science "Information Science & Library Science" and "Science and Technology Studies" subject categories are not suitable for field normalisation in bibliometric evaluations (Leydesdorff & Bornmann, 2016). A comparison of the journal level classification of publications from the Chinese Science Citation Database and the paper level classification from the Chinese Library Classification for the same dataset showed that about half of the papers could be misclassified using the journal classifications system (Shu et al., 2019), confirming that journal-based field categorisation systems could be problematic for research assessment exercises (Leydesdorff & Bornmann, 2016; Sīle et al., 2021; Wang & Waltman, 2016; Klavans & Boyack, 2017a).

Journal-based classification have problems with classifying research published in multidisciplinary journals, fast changes in research areas and making comparative analyses using different databases (Archambault, Beauchesne, & Caruso, 2011; Waltman & van Eck, 2019). For instance, using visualization and natural language processing techniques, bibliometric indicators (the h-index and the impact factor) derived from three Web of Science medical subject categories (Cardiac & cardiovascular systems, Clinical neurology, and Surgery) may provide invalid results because even within an individual subject area there could be substantial differences in terms of citation practices and impact (van Eck et al., 2013).

## 8.2. Article-based field classification systems

Theoretically, article-level classification systems can more accurately reflect scientific fields because it is common for journals to publish articles from more than one discipline, especially if they are in multidisciplinary or other generalist journals.

The academic databases PubMed, Eric, Library and Information Science Abstracts, CAB Abstracts and PsycINFO have a list of controlled vocabulary terms attributed to articles (e.g., MeSH or ERIC thesaurus)[19] and there have been attempts to use these human constructed classification schemes to improve field normalised indicators, such as with the EconLit database for economic publications (van Leeuwen & Calero Medina, 2012) or the Chemical Abstracts database (Neuhaus & Daniel, 2009; Bornmann, Marx, & Barth, 2013).

Many statistical methods, bespoke algorithms and general machine learning algorithms have been proposed to individually classify the subjects of articles based on references, citations and metadata. The disadvantages of these methods are a lack of transparency compared to journal classification systems and a lack of replicability due to changes in citation databases. For example, an early algorithm used cited references to classify articles in multidisciplinary journals (Glänzel, Schubert, & Czerwon, 1999). Some small-scale studies have shown that automatic classification is possible for some datasets. For example, one system had an accuracy of over 91% for 680 articles in 10 different topics (e.g., biotechnology, technology, fisheries, education, economics) published in a science journal by a Vietnamese university (Dien, Loc, & Thai-Nghe, 2019). Larger scale systems are essential for REF-related tasks, however.

---

[19] For other examples see https://connect.ebsco.com/s/article/Which-EBSCOhost-database-authorities-have-limited-support-via-the-EBSCOhost-API-AuthoritySearch-Method?language=en_US

Several medium scale systems have compared different approaches to classify individual fields. Four supervised machine learning algorithms were trained on article titles and abstracts of journal articles (n=66,251) indexed by the Sociological Abstracts to automatically classify sociology research into a pre-defined set of categories. The popular Gradient Boosting Classifier algorithm was shown to correctly classify over 80% of the documents (Eykens et al., 2019). A follow-up paper about Education, Economics and Sociology also used article titles and abstracts from ERIC, EconLit and Sociological Abstracts databases (113,909 records) to automatically classify articles into multiple subject categories. In this more complicated experiment, 46% of the label combinations were predicted by the Gradient Boosting model (Eykens, Guns, & Engels, 2021).

Several large-scale classification approaches have been developed for the whole of science. A multinomial logistic regression model to classify articles has been built with the help of 11 million papers from more than 4,000 journals across all academic subjects. It exploits terms in article titles, and abstracts as well as cited references. The system was subsequently used to classify 25 million articles from Scopus into four pre-defined research levels, showing that it is scalable for large problems (Boyack et al., 2014).

Another very large-scale study used deep learning applied to characters extracted from article metadata to classify scientific publications into 176 Science-Metrix subfields, and the results were compared with bibliographic coupling, direct citation, and manual-based classifications. The study used 41 million Scopus indexed publications for the experiment and found that the deep learning algorithm, despite its crude inputs, could classify scientific publications with almost the same level of accuracy as the other classification approaches (Rivest, Vignola-Gagné, & Archambault, 2021). Using a deep attentive neural network (DANN) to systematically classify publications into 104 Web of Science subject categories, another experiment was trained on 9 million abstracts from the Web of Science. The best model achieved was "micro-F1 measure of 0.76 with F1 of individual subject categories ranging from 0.50 to 0.95" (Kandimalla et al., 2021).

A recent sophisticated study used a novel unsupervised machine learning method to cluster individual publications into broad scientific disciplines and subfields. The algorithm harnessed direct citation relations between publications, clustering them into research areas and assigning labels to the research areas by extracting terms from the titles and abstracts of the publications in them. The first level of the resulting clusters included 10 to 20 major fields, the second level 500 to 1,000 fields, and the third level 20,000 to 25,000 very narrow topics (Waltman & van Eck, 2012). This method has been harnessed to find topics and specialties of pre-defined sizes (Sjögårde & Ahlgren, 2018, 2020). Using over 58 million papers, another study extended the direct citation method to create the model for classification of science into 91,726 topics that were assigned to 12 broad fields (Klavans & Boyack, 2017a). These unsupervised approaches have the advantage of probably generating finer grained and more accurate clusters of documents compared to systems that use pre-existing classes. Another advantage is flexibility to changes in research topics, but a disadvantage is a lack of transparency and the possibility that some clusters may not be meaningful groupings.

Although the accuracy of algorithms targeting the same final classification scheme can easily be compared, it is less straightforward to compare algorithms that produce different classification schemes or that cluster documents into new groups. A novel approach to solve this problem is to treat articles with many references as "gold standards" that probably broadly delineate a research area as a topic-based review (Klavans & Boyack, 2017b). Using this approach, algorithms exploiting direct citations produce better results than algorithms using bibliometric coupling (two articles citing the same paper) or co-citation (two articles cited by the same paper) and better results than the well-known journal-level classification schemes. The pre-eminence of article-level classification compared to journal level classification has been verified for human classifications too (Shu et al., 2019). The main automatic clustering approaches can be tested with free bibliometric clustering and visualisation software (Aria & Cuccurullo, 2017; van Eck & Waltman, 2017). It is not clear yet whether systems based on direct citations (i.e., clustering documents based on whether they cite each other) could be enhanced by systems that also exploit article title and abstract text.

## 8.3. Dimensions document categorisation

Dimensions is discussed separately here because it is part of a major scholarly database. It uses a machine learning algorithm applied to document titles and abstracts to automatically assign subject categories to millions of documents (e.g., papers, grants, clinical trials, patents, datasets and policy documents). Dimensions provides different schemes for categorising documents, but primarily uses the Fields of Research (FoR) categories from the Australian and New Zealand Standard Research Classification. It also incorporates other classification systems, including REF Units of Assessment (UoAs), although some are only available in subscription versions. Only the Fields of Research (FoR) and Sustainable Development Goals classification schemes are available in the free version of Dimensions[20]. For instance, Dimensions Analytics, which needs a subscription, provides a 6-digit level for categorisation of science, such as Informetrics (080705) under the broader category Library and Information Studies (0807). This could be useful when a narrow classification scheme is required for field normalised impact indicators (see Herzog & Lunn, 2018).

The Dimensions automatic field classification was initially not very accurate for library and information science (Bornmann, 2018), but its algorithm was subsequently strengthened (Herzog & Lunn, 2018). A small-scale report about a random sample of 1,000 articles from different fields compared the classification accuracy of Web of Science, Scopus and Dimensions with three independent human classifiers, finding that Web of Science had the most accurate subject classification followed by Dimensions and Scopus (Singh et al, 2020). A comparison of the classifications for research articles and letters published in Nature (2010–2020) in the Web of Science (journal level), Dimensions (machine learning) and Springer Nature (author-selected), found significant differences between paper-level classifications. Only a quarter (27%) of the papers had the same fields and 59% had partially identical subjects in Dimensions and the Springer Nature classification systems, suggesting that there are substantial differences between article-level machine classification and human-based journal approaches to classify science. More than half (52%) of the publications had identical classifications in the Web of Science and Springer Nature classification schemes and almost a third (32%) had overlapping subjects (Zhang et al., 2022b).

[20] https://app.dimensions.ai/browse/categories/publication/for

# 9. Predicting journal article citation counts or output quality from open review text

Text mining can be applied to online reviews. Increasingly many papers have open reviews within publishers' websites or public review sites (e.g., PubPeer) so this is a promising avenue for the future. It might be useful for monograph evaluation, given that reviews are important and common for monographs in the humanities (e.g., published in journals) but despite this, monograph reviews are usually positive so do not obviously form a useful source of evaluation evidence. For journal articles, open reviews are currently available for a minority of articles from publishers that permit it like MDPI. Moreover, reviews are often difficult to parse for text mining (e.g., commented copies of reviewed article PDFs). Most open reviews are from reviewers and address pre-final versions of the article, so it is not clear that they provide useful information about the final published article.

There are several open review platforms sharing pre-publication reviews (formal reviews or editorial comments from the publishing journal) or post-publication comments (recommendations or feedback by researchers or experts) for scholarly publications. ***Publons*** (publons.com) is an open review platform from Clarivate Analytics claiming to include "over 6.9-million reviews for more than 5,000 partnered journals"[21]. Partner journals can share pre-publication reviews publicly on this site, with reviewers and authors deciding what information to reveal (e.g., review text, reviewer identities). Others can also write post-publication comments on the site and anonymously score articles for significance and rigour (see: https://publons.com/benefits/reviewers/how, e.g., https://publons.com/publon/353160/). ***PubPeer*** (pubpeer.com) is another online open platform but focuses on post-publication peer review, where researchers can provide feedback or comments about published research and authors can respond. Reviews in PubPeer have identified mistakes published in leading cell biology journals [22] [23], suggesting that post-publication reviews might be a helpful source for quality assessment of published research. ***ScienceOpen*** (scienceopen.com) combines publishing and promotion services for journals with a recommendation capability where other researchers or experts can write public reviews and use a five-star score about the "importance", "validity", "comprehensibility", and "completeness" of published research[24]. ***Peer Community in*** (peercommunityin.org) is a free recommendation platform for preprints. It publishes peer-reviews of preprints in 14 subject areas, including Ecology, Genomics, Animal Science, Evolutionary Biology. The peer review process is managed by 1,700 'Recommenders' making editorial decisions about public reviews[25].

**Multidisciplinary Digital Publishing Institute** (MDPI) and several other publishers and journals provide an open peer-review option, where authors can decide to publish their reviews and reviewers can choose to be named or remained anonymous (for a review see Wolfram et al., 2020). These provide a collective source of open peer review reports for the journals covered.

In contrast to the above, **Faculty Opinions** (facultyopinions.com, previously F1000 Prime) is a paywalled source of post-publication biomedical research reviews written "by over 8,000 experts in the Life Sciences and Medicine". Articles are classified based on contribution type, such as 'Good for

[21] https://publons.freshdesk.com/support/solutions/articles/12000012231-what-is-publons-and-why-partner-with-us-
[22] https://phys.org/news/2013-05-stem-cell.html
[23] https://www.nature.com/articles/nature.2013.13060
[24] https://blog.scienceopen.com/2016/05/peer-review-at-scienceopen-is-surprisingly-simple/
[25] https://peercommunityin.org/about/

Teaching', 'New Finding', 'Technical Advance,' or 'Interesting Hypothesis' and can be given one (Good), two (Very Good) or three (Exceptional) stars[26].

The rationale behind investigating these sites is that unbiased and high-quality open reviews by subject experts might provide a further quality control mechanism for research assessment exercises, or inputs for future machine learning exercises. However, the manipulation of post-publication reviews could be problematic for formal assessments of individual academic outputs. Moreover, the current platforms for pre-publication reviews often do not show review reports for rejected articles (Thelwall, 2022a) and may have a positive bias. For example, F1000 article recommendations are exclusively positive: 'Good' (58.6%), 'Very Good' (34.6%) and 'Exceptional' (6.9%) (Waltman & Costas, 2014). Nevertheless, a study of a sample of PubPeer comments about publications found that two thirds were related to some type of misconduct (Ortega, 2022), which might be useful to flag to REF reviewers or incorporate within machine learning approaches.

The rest of this subsection reviews research into these peer review sites. Often experimental tests of site content or ratings correlate them with citation counts for the articles. Since citation counts associate with (but do not measure) research quality in many fields, the absence of research quality evidence, it is reasonable to use citation counts to help investigate the value of other indicators.

## 9.1. Publons

**Insufficient evidence of an association between Publons reviews and citations:** An early study found weak or no significant associations between bibliometric indicators from Google Scholar and the peer review activity of Publons users (Ortega, 2017) and another study of 45,819 articles from Publons also found low or insignificant correlations between bibliometric scores (e.g., WoS or Scopus citations) and Publons metrics (e.g., Quality, Significance and Overall Publons score of articles) (Ortega, 2019). These suggest that Publons metrics might not be useful indicators of citation impact or, by extension (because the two correlate in many fields), research quality.

**Articles with more positive Publons post-publication reviews receive more citations:** A small-scale paper about four experimental groups of papers from Publons with neutral, negative, positive and both negative and positive post-publication reviews found that papers with positive reviews had significantly more citations ($rho=0.498$, $p < 0.05$) while very low or non-significant associations were found between citation counts and other review polarities (Zong et al., 2020).

**Imbalances and problems in Publons content:** There has been a criticism about the coverage and contents of peer review in Publons (Teixeira da Silva & Al-Khatib, 2019; Teixeira da Silva, 2020) and a 2018 study identified a large imbalance in the coverage of Publons reviews across subjects and journals. For instance, the Publons coverage of Life Sciences (40%) was twice as much as for the Physical Sciences (18%). Most articles reviewed in Publons were from Frontiers Media open access journals (Ortega, 2019). A large-scale report about a sample of 183,743 unique review reports submitted to Publons found that although most of Publons reviews were for legitimate journals (96.7% or 177,666 for 6,403 journals), with very few reviews for apparently predatory journals (3.3% or 6,077 reviews for 1,160 journals). The share of predatory reviews was higher from sub-Saharan Africa (22%), Middle East and North Africa (14%) and South Asia (7.0%) than from other regions including North America (2.1%), Latin America and the Caribbean (2.1%), Europe and Central Asia (1.9%) and East Asia and the Pacific (1.5%) (Severin et al., 2021). There have been criticisms of the quality of post-publication reviews (e.g., da Silva & Al-Khatib, 2021, see also above for Severin et al., 2021), the recognition system for reviewing activities (Smith, 2016) and using Publons to identify potential reviewers (Jorm, 2021). Nevertheless, Publons pre-publication reviews may help universities to access formal peer review reports to "improve and promote research excellence assessment" (Wilkinson & Down, 2018).

[26] https://facultyopinions.com/wp-content/uploads/2020/07/Faculty-Opinions_Reference_Guide.pdf

**Gender gap in Publons top reviewers:** A paper about the gender of Publons top reviewers (the Top 1% most active reviewers in a field) found that male reviewers dominate across all 23 subject areas. Social Sciences, Psychiatry & Psychology and Immunology had the highest proportion of female reviewers (25%, 24.5%, 22.3%), whereas in Mathematics, Physics and most engineering subjects this proportion was less than 10%. On average, the top male Publons reviewers had significantly more reviews than the top female Publons reviewers (means: 166 vs. 117 respectively), although female reviewers tended to write longer reviews (average number of review words: 380 vs. 333). The study found low but significant Spearman correlations between reviewing activity and the number of publications of the top Publons male (0.36) and female reviewers (0.44) (Zhang et al., 2022a). In summary, Publons reviews tend to be disproportionately written by male reviewers and disproportionately cover life sciences topics.

## 9.2. Faculty Opinions (formerly F1000Prime)

**F1000Prime article factor and recommendations associate with citation counts:** Several investigations have reported significant associations between F1000Prime (now Faculty Opinions) ratings and citation metrics. An initial study of a sample of 1,397 F1000Prime selected Genomics and Genetics articles published in 2008 found low but positive Spearman correlations between F1000 article scores and citation counts from Web of Science (0.303), Scopus (0.300) and Google Scholar (0.295). Similarly, low significant correlations were found between article evaluator counts and citation counts, although a higher association was found between both F1000 article factors and citation counts (0.369) and Journal Impact Factors (0.353) (Li & Thelwall, 2012). A later study of 125 articles published in 2008 in Cell Biology or Immunology also found low-medium correlations between F1000 article factors and seven bibliometric metrics, although Percentile in Subject Area and Web of Science citation counts had higher associations with F1000 Prime article factors ($R^2$= 20% and 18% respectively) (Bornmann & Leydesdorff, 2013). Similarly, a report about a random sample of F1000 medical publications in 2007 (n=350) and 2008 (n=550) also found significant positive, albeit low, Spearman correlations between F1000 article factors and citation counts for both 2007 (rho=0.383) and 2008 (0.300), and citation counts varied between F1000Prime article type classifications, such as "new finding" and "changes clinical practice" (Mohammadi & Thelwall, 2013). Another later study also confirmed that article type has a significant effect on the difference between F1000Prime article factor scores and citations (Du, Tang, & Wu, 2016).

A large-scale study of all F1000 Prime recommendations about publications found that 2% of biomedical research had at least one F1000 Prime recommendation (WoS matched rate=93% or 95,385 publications). The study also reported weak but significant Pearson correlations between both F1000Prime recommendation scores and the number of recommendations and citation counts (0.24 and 0.26 respectively) and journal citation scores (0.33 and 0.34), suggesting that one reason for the weak associations could be that the majority (98%) of biomedical research does not have any kind of recommendation in F1000 (Waltman & Costas, 2014).

Using negative regression analysis to estimate factors influencing F1000 recommendation scores, a study of 94,641 F1000Prime publications 2000-2004 found a significant association between F1000 scores citation counts, the journal impact factor had stronger association with the citation impact of research than F1000 recommendations (Bornmann, 2015). Other studies also reported low but significant associations between F1000Prime recommendations and citation or reader metrics (e.g.,

Bornmann & Leydesdorff, 2015, Smith et al., 2019). Finally, a study of 830 research articles in 2010 published by four high impact journals found that articles recommended in Faculty Opinions (previously F1000Prime) on average had more citations (between 2010 to 2019) and this difference was statistically significant for articles published in Nature Genetics, Nature Medicine, and PLoS Biology except for the journal Cell (Wang & Su, 2021).

**Insufficient evidence of an association between F1000Prime scores and citation metrics:** An early small comparison of 1,530 publications from seven major ecological journals in 2005 and 103 publications that were highlighted by F1000 found that publications highlighted by F1000 were cited less frequently than other articles (Wardle, 2010).

**Weak associations between F1000 Prime quality assessments of biomedical publications and their research collaboration:** A study of 16,557 papers published between 1996 and 2012 found very weak associations between F1000Prime member scores (as an indication of the paper quality) and collaboration indicators. Using regression models, only 1% of the variance in the F1000 ratings was explained by the number of authors, number of affiliations, and number of countries, whereas the number of authors alone could explain 7.7% of the citation scores (Bornmann, 2017).

**Expert assessment about importance of research associated with F1000Prime ratings:** A study of 687 papers associated with the Wellcome Trust published in 2005 found a moderate positive Spearman correlation (rho=0.445) between the importance of papers, as judged by Wellcome Trust reviewers (each paper was reviewed by two reviewers using a four-point scale), and those identified by the Faculty of 1000 experts. The study also found a moderate positive correlation (0.45) between Wellcome Trust reviewer ratings and citation counts to articles three years after the reviews had been conducted. The Journal Impact Factor had the strongest association (0.625) with Wellcome Trust reviewer ratings, suggesting that journal impact might be helpful proxy indicator of research quality (Allen et al., 2009).

## 9.3. MDPI

**Disciplinary differences in characterises of standard open peer-review:** A study of 45,385 open standard article reviews for 288 MDPI journals found a large disciplinary differences in review lengths, reviewer anonymity, review outcomes, and the use of attachments. For instance, reviewers in the Physical Sciences are most likely to ask for major revisions and to use attachments in the review process, although they are less likely to disclose their identity. In Life Sciences and Social Sciences fields, reviewers tend to write longer review reports than in the Physical Sciences (Thelwall, 2022a).

# 10. Publishing quality control

Various text mining AI programs have been developed to help with quality control for papers, some of which are routinely used by publishers. Probably the most widely-used are plagiarism detection and biomedical image manipulation software. A related approach is paraphrase detection (so called "tortured phrases") which points to hidden plagiarism, sometimes of whole articles. Software has also been developed to check the plausibility of statistical results in papers by testing if the numbers reported for a test are theoretically capable of having been generated at the level of rounding reported. Whilst problematic articles seem likely to be rare in the REF, the systems represent types of approach that might support future REFs, especially if they become robust and easy to use.

## 10.1. Tools to recommend journals for manuscripts

Identifying relevant journals or conferences is the most significant step to publish research results. Hence, several publishers have developed web services to help authors to find relevant journals for manuscripts mainly based on comparing article texts (titles, abstracts or keywords) against previously published articles. Current journal recommendation tools include ***Springer Nature Journal Suggester***[27], ***Wiley Journal Finder***[28] or ***IEEE Publication Recommender***[29]. ***EndNote Manuscript Matcher*** also uses manuscript title, abstract, and references and Web of Science data to suggest related journals for manuscripts[30].***The Journal/Author Name Estimator (JANE)*** is another free service that uses the similarity scores and PubMed data to suggest most relevant journals based on manuscript titles or abstracts[31].

Various types of AI are used in these systems to match the subjects of manuscripts to related journals. For instance, ***Elsevier's JournalFinder***[32] service "uses smart search technology and field-of-research specific vocabularies" to match paper to scientific journals, ***Taylor & Francis Journal Suggester***[33] applies "artificial intelligence to match the subjects covered in articles" and ***Sage Journal Selector***[34] utilizes "an advanced AI technology" to recommend journals with similar published articles. Several studies have shown that AI or machine learning can be useful to identify appropriate academic journals or conferences with relatively high accuracy for papers (e.g., Wang et al., 2018; Feng et al., 2019; Ghosal et al., 2019a; Pradhan & Pal, 2020). For instance, a recent experiment used the XGBoost algorithm and different features (title, abstract, and keywords) from 20,250 articles from Web of Science indexed computer technology journals, reporting an accuracy of 84% for academic journal recommendations (ZhengWei et al., 2022). Using deep learning techniques, another study developed a journal recommendation system (***Pubmender***) to propose appropriate PubMed journals based upon articles' abstracts. The experiential dataset included abstracts of over 880,000 papers from 1,130 PubMed Central journals, reporting an accuracy of 87%. This was claimed to be much higher than Elsevier's Journal Finder and Springer's Journal Suggester tools (Feng et al., 2019)., The tool ***GraphConfRec*** has been developed to recommend relevant computer science conferences based on paper text, co-authorship and citation networks (Iana & Paulheim, 2021).

---

[27] https://journalsuggester.springer.com/
[28] https://journalfinder.wiley.com/search?type=match
[29] https://publication-recommender.ieee.org/home
[30] https://endnote.com/product-details/manuscript-matcher/
[31] https://jane.biosemantics.org/
[32] https://journalfinder.elsevier.com/
[33] https://authorservices.taylorandfrancis.com/publishing-your-research/choosing-a-journal/journal-suggester/
[34] https://www.edanz.com/journal-selector

The approaches used to recommend journals for manuscripts might be customised to recommend REF or journal/conference reviewers by matching the manuscripts against the publications of potential reviewers. Separate software has been generated for reviewer selection, however, as discussed below.

## 10.2. Tools to automate editorial process

AI and Natural Language Processing can assist editors and publishers in many ways from sending automatic emails to authors or reviewers and checking plagiarism in the manuscripts to statistical test and methods checking (For reviews see: Price & Flach, 2017; Checco et al., 2021; Lin et al., 2021). One study compared different manuscript management tools (e.g., ScholarOne, Editorial Manager, EVISE and Open Journal Systems) from the authors', reviewers', and editors' perspectives, reporting tools supporting automatic editorial tasks (e.g., sending e-mails, reviewer recommendation, statistical analysis, similarity check, linking references to Crossref or and PubMed) (Kim et al., 2018). A Jisc report on "*Artificial Intelligence, Automation and Peer Review*" has summarised tools that can automate some tasks of editorial management or peer review process such as the following.

- **Plagiarism detection**: *Ithenticate*[35] detects copied text in a manuscript. A procedure to detect tortured phrases – can identify plagiarism in the form of papers or sections of papers that have been processed by paraphrasing software or that have been translated from English to another language and back again. It identifies meaningless phrases that are non-idiomatic translations of scientific phrases, such as "counterfeit consciousness" from "artificial Intelligence" (Cabanac et al., 2021).
- **Automated statistical checking**: ***StatReviewer***[36] checks manuscripts against standardized reporting guidelines and *StatCheck*[37] to detect statistical errors in the submitted works (see Thelwall, 2019).
- **Multipurpose manuscript evaluation**: Other AI-assisted tools assess multiple quality control aspects of manuscripts such as *Frontiers AI Review Assistant*[38] or *UNSILO Manuscript Evaluation*[39] (see also Heaven, 2018).
- **Article summarisation**. Many natural language programs can summarise the contents of academic and other documents. UNSILO is an example.
- **Manuscript structure checking**: *Penelope.ai*[40] checks if the structure of a manuscript meets a journal's submission guidelines for the title page, abstract, citation style, references, tables and figures and information about other sections of articles (e.g., funding, acknowledgements, keywords and data/ethics statements). This avoids the need for manual checks by reviewers, publishers or editors.
- **Reference matching with in-text citations**: *Recite*[41] automatically checks and highlights if citations in the manuscript text match the reference list and vice versa.
- **Summarising articles**: *Scholarcy*[42] applies AI to summarise articles and to extract tables, figures and references.

[35] http://www.ithenticate.com
[36] http://www.statreviewer.com/
[37] http://statcheck.io/
[38] https://blog.frontiersin.org/2020/07/01/artificial-intelligence-peer-review-assistant-aira/
[39] https://unsilo.ai/unsilo-manuscript-evaluation/
[40] https://www.penelope.ai/faq
[41] https://reciteworks.com/
[42] https://www.scholarcy.com/

- **Article writing**: *SCIGen*[43] uses Natural Language Generation (NLG) to automatically generate academic papers for conferences. This has been used for quality control purposes, such as testing whether journals have genuine review procedures.
- **Methods checking**: *SciScore*[44] generates an automated assessment of articles methods on a scale of 1-10 and other reports (Design Analysis Reporting checklist and the Rigor and Transparency Index), assisting reviewers to find key information throughout a paper in a standard format (see also Menke et al., 2020).
- **Review sentiment extraction**: *PeerJudge*[45] uses AI-assisted sentiment detection to estimate the strength of praise and criticism in peer review reports on academic papers that could be useful for editorial management decisions when they want to analyse a large number or review reports. PeerJudge can predict F1000Research reviewer decisions with a moderate degree of accuracy (Thelwall et al. 2020).

[43] https://pdos.csail.mit.edu/archive/scigen/
[44] https://www.sciscore.com/
[45] http://sentistrength.wlv.ac.uk/PeerJudge.html

# 11. Allocation of research outputs to appropriate reviewers

Editorial systems for publishers suggest possible reviewers for submitted articles, perhaps based on references in the submitted outputs or by matching article keywords to the keywords of registered reviewers. REF subpanel chairs have to start the review process by assigning articles to at least two reviewers. Fully or partly automating this labour-intensive process might improve the overall match between subpanel reviewers and articles and save time.

## 11.1. Tools for reviewer selection

With the constant increase in the number of manuscript submissions to the academic journals and conferences, the editorial and peer review process is becoming more challenging and time-consuming. It was estimated that "over 15 million hours" are spent on reviewing rejected papers each year (American Journal Experts, 2018). For example, 1.2 million manuscripts are submitted to 2,300 Elsevier journals every year and only 30% (about 350,000) are published (Tedford, 2015). Due to increasing numbers of submissions and peer review workloads, a report from BioMed Central and Digital Science entitled "What might peer review look like in 2030?", recommended to "use technology to support and enhance the peer review process, including finding automated ways to identify inconsistencies that are difficult for reviewers to spot" (Burley & Moylan, 2017, p. 3). Several programs have also been developed by commercial publishers to help editors identify suitable reviewers (see examples below), but these are typically not transparent.

- **Clarivate's Reviewer Locator**[46] automatically suggests reviewers based on data from the Web of Science and Publons peer review databases and connects to the ScholarOne submission management system integrating editorial and peer-review processes.
- **Reviewer Discovery**[47] is a tool from Aries Systems that uses ProQuest author profiles and automatically suggests reviewers based on the Editorial Manager system.
- **Elsevier's EVISE**[48] uses Reviewer Finder to identify and recommend reviewers based on Scopus data.
- **Frontiers Coronavirus Reviewer Recommender**[49] suggests experts to review COVID-19 research proposals using "Frontiers knowledge graph and advanced information extraction and retrieval methods".

The National Natural Science Foundation of China (NSFC) has developed an AI-assisted reviewer recommender for grant applications using natural language processing and an assignment decision support system to help select expert panels. An initial version of the AI system had chosen "at least one member of each of nearly 44,000 panels that approved projects" in 2018, and the accuracy of system was about 80% (Cyranoski, 2019, p. 317), but accuracy improvements are still being made (Liu et al., 2022). The system classifies the reviewers and proposals by discipline and uses information from scientific databases (e.g., Web of Science) and referee profiles in NSFC databases about the publication records or research projects of potential reviewers and then uses lexical semantic analysis to compare the extracted information with the grant applications. Different rules were used in the system to avoid conflicts of interests between reviewers and applicants (e.g., affiliation, co-authorship, project and tutor-student relationships) (Liu et al., 2016). The Toronto Paper Matching System also automatically suggested reviewer assignments for the NIPS 2010 conference using a topic modelling approach to estimate reviewers' expertise areas. The system extracts publication records from Google Scholar to generate profiles for reviewers and uses supervised score prediction models to suggest reviewer assignments (Charlin & Zemel, 2013). Several other studies have also suggested algorithms for the automatic assignment of reviewers to conference papers (e.g., Li & Watanabe, 2013; Al Mahmud, Hossain, & Ara, 2018; Kalmukov, 2020), mostly for AI related conferences. No robust accuracy measures seem to have been generated for these systems yet, however. Presumably the ground truth for such a system would be human editor assignments or (in conferences that allow this) reviewer requests to review. **The implementation of such systems seems to be practical and beneficial for the REF.**

[46] https://clarivate.com/webofsciencegroup/solutions/reviewerlocator/
[47] https://www.ariessys.com/newsletter/june-2013/reviewer-discovery-free-trial-available/
[48] https://www.elsevier.com/connect/reviewers-update/rolling-out-our-new-editorial-system-evise
[49] https://coronavirus.frontiersin.org/covid-19-research-funding-monitor

Two topics are discussed in more detail below. These are relevant to the REF and have academic research about them.

**Review decision and comment automation** is challenging with limited progress on a few topics so far. Whilst positive correlations between human and automated decisions have been generated, no current system challenges human reviewing yet. Positive correlations between peer review judgements and machine learning do not necessarily mean that further progress is likely soon because an AI system would achieve a positive correlation by rejecting papers with obvious errors, such as very poor grammar, too short, or lacking references. ReviewAdvisor[50] is a natural language processing toolkit designed to help select good manuscripts for a journal and provide feedback to help authors improve their submitted articles. Whilst its performance on the authors' ASAP-review set of 28,119 machine learning conference paper reviews was weak, it provides a starting point and might help reviewers by suggesting comments on paper aspects that they may have overlooked. Another study developed an AI tool using trained neural network and a set of features from papers including word frequencies, readability scores, and formatting measures, finding that automated systems developed unethical biases, such as against grammar and formatting errors, that helped them be more accurate (Checco et al., 2021). Similarly, another study found that Natural Language Processing models to generate reviews for scientific papers could make the peer-review task easier and more effective but not to replace it (Yuan, Liu, & Neubig, 2021).

**Review decisions from peer review comments:** Based on dataset of scientific peer reviews from PeerRead[51], a deep learning network was used to predict acceptance or rejection of articles from peer review reports and to generate the final meta-review, finding "good consistency between the recommended decisions and original decisions", with 74%-86% accuracy at predicting the binary decision accept/reject, which was better than standard machine learning algorithms and prior bespoke peer review judgement algorithms (Pradhan et al., 2021, p. 237). Another study used 3,341 papers from three computing conferences (2018 IEEE Wireless Communications and Networking Conference and 2018 and 2019 the International Conference on Learning Representations) and assessed to what extent AI can predict human peer-review decisions about papers (acceptance/rejection or average review score). Using a Random Forest classifier, the above study used machine learning to predict the acceptance of papers submitted to a top AI conference with 81% accuracy based on surface features, such as the number of tables in the papers or characters in the title (Skorikov & Momen, 2020). Similarly, the pReview software package developed for automatically generating summarization, contribution detection, writing quality analysis and potential related works of academic papers to support reviewers (Bond & Craig). There is also evidence that sentiment analysis of review reports could be helpful to predict the final decision (acceptance or rejection) of conference papers (Wang & Wan, 2018; Ghosal et al., 2019b; Chakraborty, Goyal & Mukherjee, 2020) or review scores of funding programs (Luo et al., 2021).

[50] https://github.com/neulab/ReviewAdvisor

[51] https://paperswithcode.com/paper/a-dataset-of-peer-reviews-peerread-collection

## 12. Conclusions and recommendations

Many bibliometric studies have shown that a wide range of factors derived from article text (e.g., length of articles, titles or abstracts, number or impact of cited references and article readability) might be related to the scientific impact of research as measured by citation counts (see Table 1 for summary). However, there are disciplinary differences in almost all the results and some findings could be biased towards impactful research. Some of the results also varied over time or between journals. Thus, whilst there are general trends, there are no universal laws. A risk with text mining to predict citation counts is that it is likely to work best by identifying highly cited topics, predicting higher citation counts for all articles on these topics. A successful prediction model for one year might be invalid for the next one due to topic changes, so text mining may need rebuilding each year to identify the new hot topics.

Many studies have also found that several factors derived from article metadata (e.g., the number of authors, institutional or international collaboration, journal impact or author publication and citedness) sometimes associate with the citation rates of published papers (see Table 2 for a summary). Nevertheless, there are disciplinary differences in some of the results and other factors such as the selection method for journals (e.g., high impact journals) or publication years might have influenced the results. It seems that the journal impact factor had the strongest association with citation counts, as reported by several studies, although collaboration factors also have strong associations with the citation impact of research. In particular, the expected citation rate of UK authored research increases roughly in line with the logarithm of the number of authors. Several machine learning studies suggested that it is feasible to predict the long-term citation counts or quality scores of papers in some extend. Some studies have used machine learning to estimate future citations for articles, although there is no evidence how accurately AI can be used to predict human judgements of journal article in a large-scale.

There is evidence that public datasets of citations (e.g., Dimensions and OpenCitations) can be useful sources to automatically capture wider citations from larger scholarly publications, especially for the recently published articles for the future REF in addition to traditional citation indexes (e.g., Scopus and Web of Science). Many studies have also shown that altermatic indicators and particularly Mendeley readers can be useful to predict future scholarly impact. However, these sources lack quality control mechanism for the future REF and can potentially be spammed such as by publishing non-peer reviewed publications online to inflate citations to the REF submissions or adding readers or bookmarks via fake accounts or manipulated users. There are also arguments about the accuracy of automatic document classification systems to assign REF outputs to the scientific fields (e.g., Dimensions).

Pre-publication open reviews platforms such as Publons and MDPI could be helpful counselling source for the future REF, when text mining and AI-assisted sentiment detection (e.g., PeerJudge) can automatically be applied to estimate the strength of praise and criticism in formal peer review reports on REF submissions. However, few journals or publishers allow public sharing of pre-publication reviews, many reviews may not have all comments from reviewers in a structured format and currently review reports for rejected manuscripts submitted to journals are not usually available. Post-publishing peer reviews (e.g., Faculty Opinions) can potentially be helpful advisory source when feedback or comments by experts in the fields about published articles could be captured, although post-publication reviews could be manipulated and hence problematic to be used.

There are several AI-assisted tools and software that might be consulted to automatically suggest reviewers (REF subpanel members) for submitted outputs, enhancing the peer review process by assigning relevant reviewers in different the subject area. However, the accuracy of the developed tools to assign reviewers is not fully known and it is not clear how they may work based on few referee profiles in some REF subjects.

Finally, the review of the literature on the responsible use of technology to assist research assessment suggested that most AI and machine learning methods have recently developed and tested and there is evidence that AI adoption continues to grow in many areas (see The State of AI in 2021[52]). Hence, the use technology to assist or even enhance evaluations of individual articles could be possible in the future.

The main recommendations are listed in the executive summary.

## Acknowledgements

Thank you to members of the steering group for comments on earlier drafts: Andy Hepburn (Research England), Steven Hill (Research England), Petr Knoth (Open University), Duncan Shermer, (Research England), and Jennifer Stergiou (University of Northumbria and ARMA Chair). This study was funded by Research England, Scottish Funding Council, Higher Education Funding Council for Wales, and Department for the Economy, Northern Ireland as part of the Future Research Assessment Programme (https://www.jisc.ac.uk/future-research-assessment-programme). The content is solely the responsibility of the authors and does not necessarily represent the official views of the funders.

[52] https://www.mckinsey.com/business-functions/quantumblack/our-insights/global-survey-the-state-of-ai-in-2021